\documentclass[useAMS,usenatbib]{mn2e}
\usepackage[dvipdfmx]{graphicx}
\usepackage{epsfig}
\usepackage[usenames]{color}
\usepackage[dvipdfmx]{pdflscape}
\usepackage{multicol}
\definecolor{orange}{rgb}{1.0,0.5,0.0}
\definecolor{chocolate}{rgb}{1.0,0.5,0.15}
\definecolor{purple}{rgb}{0.67,0.13,1.}
\title[Testing models for Long Secondary Periods]
{Using broadband photometry to examine the nature of
Long Secondary Periods in red giants}
\author[M. Takayama, P. R. Wood and Y. Ita]{M. Takayama$^{1}$\thanks{E-mail:
m.takayama@astr.tohoku.ac.jp },
P. R. Wood$^{2}$ and Y. Ita$^{1}$\\
$^{1}$Astronomical Institute, Graduate School of Science, Tohoku University, Sendai, Miyagi 980-8578, Japan\\
$^{2}$Australian National University, Research School of Astronomy and Astrophysics, 
Cotter Road, Weston Creek ACT, 2611 Australia}
\begin{document}

\date{Accepted 1988 December 15. Received 1988 December 14; in original form 1988 October 11}

\pagerange{\pageref{firstpage}--\pageref{lastpage}} \pubyear{2002}

\maketitle

\label{firstpage}

\begin{abstract}
Long-term $JHK$ light curves have recently become available for large
numbers of the more luminous stars in the SMC.  We have used these
$JHK$ light curves, along with OGLE $V$ and $I$ light curves, to examine the
variability of a sample of luminous red giants in the SMC which show
prominent long secondary periods (LSPs).  The origin of the LSPs is
currently unknown. In oxygen-rich stars, we found
that while most broad band colours (e.g. $V-I$) get redder when an
oxygen-rich star dims during its LSP cycle, the $J$-$K$ colour barely
changes and sometimes becomes bluer.  We interpret the $J$-$K$ colour
changes as being due to increasing water vapour absorption during
declining light caused by the development a layer of dense cool gas
above the photosphere.  This result and previous observations which
indicate the development of a chromosphere between minimum to maximum
light suggest that the LSP phenomenon is associated with the ejection of
matter from the stellar photosphere near the beginning of light
decline.  We explore the possibility that broadband
light variations from the optical to the near-IR regions can be
explained by either dust absorption by ejected matter or large
spots on a rotating stellar surface.  However, neither model is
capable of explaining the observed light variations in a variety of
colour-magnitude diagrams.  We conclude that some other mechanism is
responsible for the light variations associated with LSPs in red giants.

\end{abstract}


\section{Introduction}

All luminous red giants appear to be variables and they are generally
known as Long Period Variables(LPVs). They vary with periods which
fall on at least eight period-luminosity sequences which have been
labeled A$^{\prime}$, A, B, C$^{\prime}$, C, D, E and F (\citealt{woo99}; \citealt{sos04};
\citealt{ita04}; \citealt{tab10}; \citealt{sos13b}). Sequences A$^{\prime}$-C
and F are thought to consist of stars pulsating in the radial
fundamental and overtone modes and the non-radial dipole and
quadrupole modes (\citealt{woo99}; \citealt{ita04};
\citealt{takayama13}; \citealt{sos13b}; \citealt{ste14}). These stars have been
classified as Miras, Semi-Regulars(SRs) and OGLE Small Amplitude Red
Giants (OSARGs) (\citealt{woo99}; \citealt{sos04};
\citealt{sos07}). On the other hand, sequence E is known to consist of
close binary system showing ellipsoidal and eclipsing variability
(\citealt{woo99}; \citealt{der06}; \citealt{sos07b}).

The last sequence, called sequence D, exhibits the longest periods of
all the sequences.  Sequence D stars show photometric variations with
periods of $\sim$400--1500 days and they also show another variations with a
shorter period that usually correspond to the period of sequence B.
In these stars, the shorter period is called the
primary period and the longer period corresponding to sequence D is known as
a Long Secondary Period (LSP). LSP stars have been found within the
LMC, SMC and Galactic populations of LPVs.  In fact, approximately
25-50$\%$ of luminous red giants have been found to show LSPs in MACHO
and OGLE data (\citealt{woo99}; \citealt{sos07};
\citealt{fra08}). However, the origin of the LSP phenomenon is still
unknown and many explanations have been suggested.  Some of these are
now highlighted.

The most straightforward explanation is pulsation.
However, all pulsation explanations have problems.  Since, at the given
luminosity, the typical LSP is approximately 4 times
longer than the period of a Mira (on sequence C), and since the Miras
are known to be pulsating in the radial fundamental mode, an
explanation in terms of normal-mode radial pulsation is not possible.
Non-radial g-mode pulsation was proposed as the explanation
of the LSP phenomenon by \citet{woo99} since g-mode periods can be longer than
the radial fundamental mode period. However, because red giants have
a large convective envelope and only a thin outer radiative layer where
g-modes could develop, the amplitude of g-modes should be too
small to correspond to the observed light and radial velocity amplitudes of LSP
stars \citep{woo04}.

Some researchers have also argued for a close binary hypothesis (like
sequence E) to explain the LSP phenomenon. \citet{sos07b} found at
least 5$\%$ of LSP stars showed ellipsoidal or eclipse-like
variability.  He also noted, as did \citet{der06}, that if the
sequence E ellipsoidal variable were plotted using the orbital period
rather than the photometric period (half the orbital period) then
sequences D and E overlapped on the period-luminosity
diagram. However, the large amplitude of many LSP light curves (up to
1 magnitude) would require a close pair of eclipsing red giants,
very unlikely in itself, and this would lead to alternating deep and
shallow minima which are not seen.  On the other hand, \citet{woo99}
found that many light curves due to the LSPs show faster decline and
slower rise and they proposed that a binary with a companion
surrounded by a comet-shaped cloud of gas and dust could give rise to
those light curve features. Various observational and numerical
results have conflicted with the binary hypothesis. \citet{hi02}
observed radial velocity variations in six local red giants with LSPs
and found that the velocity curves all had a similar shape.  For
eccentric binaries with a random longitude of periastron, the light
curve shape should vary from star to star.  The lack of such
variability led \citet{hi02} to reject the binary explanation for
LSPs.  \citet{woo04} and \citet{nic09} obtained radial velocity curves
for stars in the Galaxy and LMC and they argued that the small typical
full velocity amplitude of $\sim$3.5 km s$^{-1}$ meant that all binary
companions would need a mass near 0.09\,${\rm M}_{\odot}$.  Since
companions of this mass are rare around main-sequence stars, it seems
unlikely that the 25--50\% of red giants which show LSPs could have a
companion of this mass.

Star spots due to magnetic activity on AGB stars have been proposed by
some researchers (e.g. \citealt{sok99}) and a rotating spotted star
has been suggested as an explanation of the LSPs.  Possible evidence
for magnetic activity in stars with LSPs is provided by the existence
of H$\alpha$ absorption lines whose strength varies with the LSP
\citep{woo04}.  However, from observations, \citet{oli03} found that
the rotation velocity of LSP stars is typically less than 3 km/s and
they noted that this rotation velocity is too small to explain the
periods of less than 1000 days which are often observed for the LSP.  For
example, a 3 km/s rotation velocity gives rise to the period of 2868
days for an AGB star with a $\sim$170 $R_{\odot}$.  A way to overcome
this objection would be to have multiple star spots.  In another
study, \citet{woo04} found that with a simple blackbody model, it was
difficult for a rotating spotted star to simultaneously reproduce both
the amplitude and colour variation observed in the MACHO observations
of stars with LSPs.  The use of model atmospheres rather than
blackbodies, and allowing for limb darkening, could alter this
conclusion.

The giant convective cell model is a relatively new explanation of the
LSP phenomenon \citep{sto10}.  This model is somewhat similar to the
rotating spot model in that it involves variable amounts of the
visible stellar surface having a temperature cooler than normal.  The
photometric variations seen in both the rotating spot model and the
giant convective cell model should be similar.

Another potential origin for the LSP is a variable amount of
circumstellar dust. \citet{woo99} proposed that the periodic dust
formation found in the circumstellar regions of theoretical models of
AGB stars by \citet{win94} and \citet{hof95} could give rise to the
light variation of the LSPs.  Evidence for dust around red giants with
LSPs was found by \citet{woo09} who showed that LSP stars tend to have
a mid-infrared excess.  On near- to mid-infrared color-magnitude
diagrams, the LSP stars lie in a region similar to that occupied by R
Coronae Borealis stars which are known to be surrounded by patchy
circumstellar dust clouds. This result suggests that the variation of
LSP stars might be due to the ejection of dust shells that may be
patchy.  Similarly, in studying the effect of dust ejection on MACHO
colours and magnitudes, \citet{woo04} noted that patchy dust clouds
seemed to be required in order to reproduce the observed
colour-magnitude variations.

In this paper, we utilize new long-term near-infrared $JHK_{s}$ light
curves of stars in the SMC obtained with the SIRIUS camera on the IRSF
1.4 m telescope \citep{ita15}. We combine the $JHK_{s}$ observations
with $V$ and $I$ monitoring of SMC stars by the Optical Gravitational
Lensing Experiment (OGLE) \citep{sos11}.  Using this combined data, we
examine light and colour variations of LSP stars over a wide
wavelength range.  We then test models involving dust ejection or
variable spots to see if they can reproduce the observed light and
colour variations over the wide wavelength range.  For our
computations, the \textsc{dusty} code (\citealt{ive99}) is used to
model the effects of ejected dust and the 2012 version of the
Wilson-Devinney code \citep{wil71} is used to model the effects on the
light and colour of a spotted star.

\section[]{The observational data}

\subsection{The near-IR data}

A long-term multi-band near-infrared photometric survey for variable
stars in the Large and Small Magellanic Clouds has been carried out at
the South African Astronomical Observatory at Sutherland
\citep{ita15}.  The SIRIUS camera attached to the IRSF 1.4 m telescope
was used for this survey and more than 10 years of observations in the
near-IR bands $J$(1.25$\mu$m), $H$(1.63$\mu$m) and $K_{s}$(2.14$\mu$m)
band were obtained. In this work, we select the SMC stars from the
SIRIUS database. Variable and non-variable stars in an area of 1
square degree in the central part of the SMC have been observed about
99-126 times in the period 2001--2012 and a total of 340147, 301841
and 215463 sources were identified in $J$, $H$ and $K_s$,
respectively. We note that the photometric detection range of the
SIRIUS camera is about 8$\sim$18 magnitudes in K$_{s}$ band.
A total of 12008 variable sources of all kinds were detected
of which 4533 were detected in all three wavebands (data release ver 130501. \citet{ita15}).

\subsection{The optical data}

We obtained $V$ and $I$ band time series of SMC red giants from the
OGLE project \citep{sos11}. Typically, about 1000 observing points
were obtained in the $I$ band by OGLE-II+III while a much smaller
number of about 50--70 points were obtained in the $V$ band.

\citet{sos11} divided the LPVs into LSP and non-LSP stars according to
their positions in the period-luminosity relations for variable red
giants.  We found nearly 700 LSP stars in the area of the IRSF
infrared survey.  In order to determine if an LSP star is oxygen-rich or
carbon-rich, we initially adopted the classification method introduced
by \citet{sos11}.  The stars we examined in detail had their
oxygen-rich or carbon-rich status checked against optical spectral
classifications in the literature.

\subsection{The combined near-IR and optical data}

By combining the SIRIUS data with the OGLE data, we obtained 298
oxygen-rich LSP stars and 88 carbon-rich LSP stars in the region
monitored by the SIRIUS camera. For all sample stars, a small number
of apparent bad data points were removed from each time series. Then,
by using a first order
Fourier fit to the $I$ band time series, we
obtained the period and amplitude corresponding to the largest
amplitude mode of the light curves. In all cases, the obtained periods
and amplitudes were almost identical to those in the OGLE-III catalog
(\citealt{sos11}).  Using the period from the fit to the $I$ light
curve, a fit was also made to the $K_s$ light curve in order to derive
its amplitude. 

\begin{figure}
\includegraphics[width=0.5\textwidth]{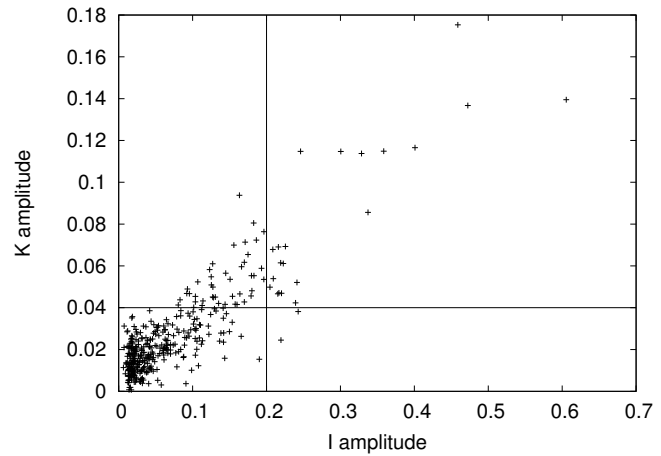}
\caption{The relation between the $I$ and $K_s$ band full amplitude of 298
oxygen-rich stars and 88 carbon stars. The stars selected for
detailed study have $I$ amplitude greater than 0.2 magnitudes and $K_{s}$
amplitude greater then 0.04 magnitudes.}
\label{Iamp_Kamp}
\end{figure}

Since our aim is to look for and to model colour and magnitude changes,
we need the observed changes to be significantly larger
than the noise in the data.  The limiting factor in this study is
noise in the near-IR data. We concentrate on the brightest
stars with $K_s < 13$ and for them the useful lower limit for 
clearly apparent $K_s$ band variation is about 0.04 magnitudes.  
Similarly, we set a useful lower limit for $I$ band variation of about 
0.2 magnitudes. Figure\,\ref{Iamp_Kamp}
shows the full amplitude of the observed LSP
stars in the $I$ and $K_{s}$ bands.  This figure shows the selected stars,
which having full
amplitudes larger than 0.04 magnitudes in $K_s$ and 0.2 magnitudes in $I$.  
This selection gave us a sample of 7 oxygen-rich stars and 
14 carbon stars for analysis.  
The spectral types of our sample stars obtained from
the OGLE photometry agree with the classifications in the SIMBAD
astronomical database where the spectral types
were obtained from spectroscopic determinations. The properties of these stars
at maximum light are listed in Table~\ref{tab01}.

Near infrared $JHK_{s}$ photometric data as a function of Julian date (JD) for a sample of 7 oxygen-rich stars and 14 carbon stars are published along with this paper. The file name of the time-series data indicates star-name and waveband, like "OGLE-SMC-LPV-07852.J.dat". Table~\ref{tab_time_series} shows time-series for a sample of the LSP star, which consists of seven columns as follows.\\
Column 1 : Julius day.\\
Column 2 : Calibrated (2MASS referenced) magnitude.\\
Column 3 : Error in deferential image analysis in magnitude.\\
Column 4 : Error in reference magnitude.\\
Column 5 : Error in conversion offset in magnitude.\\
Column 6 : Name of the survey region in which the variable star is detected.\\
Column 7 : Name of the survey subregion in which the variable star is detected.\\
for more detail, see \citet{ita15}.

\begin{table*}
\begin{center}
\begin{minipage}{180mm}
\caption{
Basic parameters of the program stars at maximum light 
}
\begin{tabular}{ccccccccccccc}
\hline
\hline
   Name & $RA$ & $Dec$ &$V$ & $I$ & $J$ & $H$ & $K_{s}$ & $L$ &
$T_{\rm BB}$ & $T_{\rm eff}$ & $P_{\rm LSP}$ & Sp.\\
  & & & (mag) & (mag) & (mag) & (mag) & (mag) & (${\rm L}_{\odot}$) &
(K) & (K) & (days) & type\\
\hline
OGLE-SMC-LPV-07852 &12.402540 &-72.97075&16.60 &14.34 &12.91 &12.05 &11.73 &4617 &2803 &3434& 645 &O \\
OGLE-SMC-LPV-09856 &13.186620 &-72.95178&16.49 &14.37 &12.96 &12.10 &11.86 &4408 &2928 &3558& 615 &O \\
OGLE-SMC-LPV-10774 &13.530120 &-72.47989& -- & 14.98 &13.73 &12.81 &12.57 &2170 &2844 &3474& 396 &O \\
OGLE-SMC-LPV-12332 &14.166840 &-73.07389& -- & 14.26 &12.95 &12.08 &11.82 &4455 &2887 &3517 &632 &O \\
OGLE-SMC-LPV-14045 &14.987085 &-72.94936& 17.50 & 15.13 & 13.72 & 12.79 & 12.57 &2197 &2865 &3495 &390 &O \\
OGLE-SMC-LPV-14715 &15.332205 &-72.86369& -- &14.27 &12.92 &12.03 &11.75 &4544 &2815 &3445 &601 &O \\
OGLE-SMC-LPV-14781 &15.361920 &-72.85678&16.48 & 14.62 &13.30 &12.44 &12.23 &3245 &2991 &3620 &469 &O \\
OGLE-SMC-LPV-08199 &12.555540 &-73.18697&17.75 &14.90 &13.29 &12.27 &11.64 &3745 &2203 &2792 &767 &C \\
OGLE-SMC-LPV-08280 &12.583995 &-72.80533&16.36 &13.78 &12.33 &11.36 &10.93 &7922 &2490 &3106 &1008 &C \\
OGLE-SMC-LPV-09341 &12.990495 &-73.20256& -- & 13.92 &12.48 &11.63 & 11.26 &6419 &2734 &3362 &1081 &C \\
OGLE-SMC-LPV-10109 &13.281915 &-73.12992&16.08 &13.84 &12.52 &11.60 &11.20 &6411 &2592 &3215 &964 &C \\
OGLE-SMC-LPV-12653 &14.317380 &-72.80128& -- &14.83& 13.10 &12.03 &11.44 &4465 &2200 &2789 &1356 &C \\
OGLE-SMC-LPV-13340 &14.654580 &-72.44625&16.28 & 14.12& 12.59& 11.66&11.28 &6003 &2606 &3230 &1014 &C \\
OGLE-SMC-LPV-13557 &14.755710 &-72.46908&16.31&13.91&12.45&11.54&11.13 &6876 &2592 &3215 &1032 &C \\
OGLE-SMC-LPV-13748 &14.847210 &-73.04900& -- &14.69 &13.09 &12.11 &11.67 &3977 &2468 &3082 &925 &C \\
OGLE-SMC-LPV-13802 &14.871165 &-72.65750&17.08 &14.19 &12.69 &11.71 &11.16 &6085 &2334 &2936 &1011 &C \\
OGLE-SMC-LPV-13945 &14.946045 &-72.71094& -- &14.06 & 12.63 &11.73 &11.29 &5872 &2560 &3180 &927 &C \\
OGLE-SMC-LPV-14084 &15.004245 &-72.89839&16.30 &14.27 &12.74 &11.83 &11.53 &5042 &2759 &3388 &799 &C \\
OGLE-SMC-LPV-14159 &15.048705 &-72.66822& -- &14.42 &12.96 &12.04 &11.57 &4406 &2495 &3112 &913 &C \\
OGLE-SMC-LPV-14829 &15.382875 &-73.09919&15.69 &13.70 &12.37 & 11.54 & 11.21 &6950 &2839 &3469 &780 &C \\
OGLE-SMC-LPV-14912 &15.434040 &-73.29506& -- &14.49 &12.92 &11.91
&11.31 &5135 &2244 &2837 &722 &C \\
\hline
\end{tabular}
Notes: $T_{\rm BB}$ is the blackbody temperature derived from $J-K_s$ while
$T_{\rm eff}$ is the effective temperature derived from $J-K_s$ 
using the formula in \citet{bes83}.  Luminosities $L$ were derived by applying
a bolometric correction to $K_s$ (see Section~\ref{sec:model}).
\label{tab01}
\end{minipage}
\end{center}
\end{table*}

\begin{table*}
\begin{minipage}{180mm}
\begin{center}
\caption{
The first ten records from the time series data for a variable source in $J$ band (OGLE-SMC-LPV-07852). The seven columns are explained in the text. The full version of this table is available as online material. 
}
\begin{tabular}{ccccccc}
\hline
\hline
Column 1 & Column 2 & Column 3 & Column 4 & Column 5 & Column 6 & Column 7\\
(day) & (mag) & (mag) & (mag) & (mag) &  & \\
\hline
2452092.635160 & 12.930 & 0.003 & 0.019 & 0.003 & SMC0050-7250 & C\\
2452212.344921 & 12.959 & 0.002 & 0.019 & 0.003 & SMC0050-7250 & C\\
2452213.302422 & 12.958 & 0.003 & 0.019 & 0.003 & SMC0050-7250 & C\\
2452214.397972 & 12.940 & 0.002 & 0.019 & 0.003 & SMC0050-7250 & C\\
2452246.289626 & 12.991 & 0.003 & 0.019 & 0.003 & SMC0050-7250 & C\\
2452285.324709 & 13.007 & 0.003 & 0.019 & 0.003 & SMC0050-7250 & C\\
2452298.305372 & 12.974 & 0.003 & 0.019 & 0.003 & SMC0050-7250 & C\\
2452390.680728 & 12.931 & 0.002 & 0.019 & 0.003 & SMC0050-7250 & C\\
2452411.689245 & 12.948 & 0.002 & 0.019 & 0.003 & SMC0050-7250 & C\\
2452415.634797 & 12.965 & 0.003 & 0.019 & 0.003 & SMC0050-7250 & C\\
\hline
\end{tabular}
\label{tab_time_series}
\end{center}
\end{minipage}
\end{table*}

\section{Separating the LSP variation from the primary period variation}

Light curves of all LSP stars show variations with the period
corresponding to sequence D and also variations with the primary
period, which mainly corresponds to sequence B. Therefore, in order to
study the LSP variation only, we need to separate each light curve into the
components due to variation with the LSP and variation with the primary
period. The separation method we used is now described.

\begin{figure}
\includegraphics[width=0.5\textwidth]{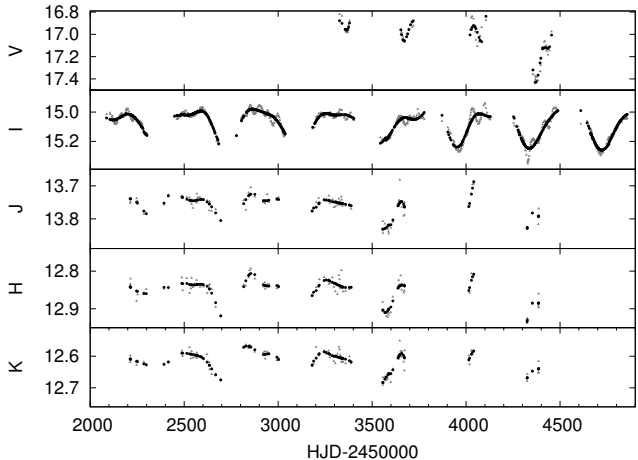}
\caption{The complete light curves of OGLE-SMC-LPV-10774 ($P$=396\,d) 
in the bands $VIJHK$.  The raw data points
are shown as small filled grey triangles while the fit values
are shown as large filled black circles.}
\label{lcfits}
\end{figure}

At first, we adopted a reference time $t_{0}$ and 
then delimited the light curve into time intervals equal to the
size of the LSP ($P$), with $t_0$ as the start of an interval. Then 
we fitted the observed data points in each interval independently 
using a second
order Fourier series $F_{0}$ with the period equal to that of the LSP,  
\[F_{0}(t)=A_{0}+\sum_{i=1}^{2} a_{0}\sin \left( 2\pi i \frac{t-t_{0}}{P}
\right) +b_{0}\cos \left( 2\pi i \frac{t-t_{0}}{P} \right).\]
The fit was performed on independent intervals of length $P$ 
rather than the full light curve interval
because the light curves of LSP stars vary substantially from cycle to cycle
of the LSP.  Furthermore, in order to minimize the dependence of the final fit on the
adopted value of $t_0$, the fit was made using 4 values  
for $t_0$, $\tau_0$, $\tau_1$, $\tau_2$ and $\tau_3$, 
with each of these values increasing by $\frac{1}{4}P$,
giving fits $F_0(t)$, $F_1(t)$, $F_2(t)$ and $F_3(t)$, respectively. 
In this study, $\tau_0$ corresponds to JD2450000.
A requirement for a fit to be made was that there be at least 6
points in each of the first and second halves of the fit interval.
If such points were not available, no fit was made.
The final adopted fit $F(t)$ at the observed times is given by 
\[F(t)=\frac{1}{4} \sum^{3}_{j=0} F_{j}(t).\]
When there were fewer than 4 fits, the average was taken over
the number of fits available.
An example of the final light curve fits for each band in a
representative star is shown in Figure\,\ref{lcfits}.

As one can see from Figure\,\ref{lcfits}, observation times of the OGLE $I$
band usually differed from those of the OGLE $V$ band and the SIRIUS $J$, $H$ and $K_s$
bands, the latter being taken simultaneously. The $I$ band observations 
are always the most frequent.  In our analysis, we will examine colours
relative to the $I$ band magnitude.  In order to
derive the colours and $I$ band magnitudes at simultaneous dates, 
the $I$ band fits to the LSP variations were used.  Magnitudes and colours
were derived from the fits at observation times of the $V$
band and the $K_{s}$ band.  As noted above, fits are not available at all
times in a given band.

\section{Modeling the LSP magnitude and colour variations}

We will now create models for the magnitude variations in various
bands associated with LSPs in order to see if the assumed models 
are consistent with the observations.  The two models we examine
are obscuration by dust and the variation of spots on a star.

\subsection{The dust model}\label{sec:model}

One possible explanation of the LSP phenomenon is episodic circumstellar dust absorption.
Here we assume that the central star ejects mass for several hundred days
in each LSP cycle, leading to the formation of thin spherical mass shells. Dust grains
which form in the ejected matter will be carried along with the expanding mass shells
leading to dimming of the light from the central star.
We assume that the dust shells are ejected spherically although this is
not an essential aspect of the dust model for the optical
and near-infrared bands which we are examining since thermal emission in
these bands is not strong. 

The amount of dimming of a star
due to circumstellar dust depends on the optical depth through the dust.
For a spherical dust shell
with an outer radius $R_{\rm out}$ (corresponding to the start of a mass
ejection episode), an inner radius $R_{\rm in}$ (corresponding to the 
end of a mass ejection episode, or the dust formation radius in a newly-forming
inner shell), a constant expansion velocity $v$
and a mass ejection rate $\dot{M}$, the optical depth $\tau_{\lambda}$ through the shell
is given by \citep[e.g.][]{gro93}
\begin{equation}
\tau_{\lambda} \propto \frac{\dot{M}\Psi Q_{\lambda}/a}{v \rho_{\rm g}} (\frac{1}{R_{\rm in}} - \frac{1}{R_{\rm out}}), 
\label{opt_depth}
\end{equation}
where $\Psi$ is the dust to gas mass ratio, $Q_{\lambda}$ is the
absorption efficiency of the grain at wavelength
$\lambda$, $a$ is the grain radius and $\rho_{\rm
  g}$ is the grain density.

We have used the \textsc{dusty} code
(\citealt{ive99}) to examine the effect of a spherical dust shell on
the broad band colours of LSP stars. 
\textsc{dusty} computes the SED of a dust enshrouded LSP star
for one given configuration of the dust shell.  
From Equation~\ref{opt_depth}, we can see that the innermost dust
shell plays the most important role for dust extinction 
when multiple dust shells of similar origin exist.  In our
model we only consider the innermost dust shell.

The first input parameter for \textsc{dusty} which we consider is the dust shell thickness.
AGB stars are well-known mass-losing stars in which the
terminal velocity of the mass flow is
$\sim$10--20 km s$^{-1}$ (\citealt{dec07}; \citealt{deb10};
\citealt{lom13}).  On the other hand, R
Coronae Borealis (RCB) stars are similarly luminous evolved stars which
eject mass semi-periodically in random directions (\citealt{cla96}). 
The large-amplitude variability of these stars is due to 
dust absorption by those ejected
dust clouds which lie in the line of sight to the star. Observationally, the
ejection velocity of the clouds is $\sim$200--250
km s$^{-1}$ (\citealt{fea75}; \citealt{cla96}; \citealt{fea90}).  
In view of these velocity estimates, and since we
do not know what causes dust shell ejections in LSP stars, 
we considered two values for the wind velocity $v$ of
15 and 220 km s$^{-1}$, and the velocity is assumed to be constant with
radius.  We assume that dust is ejected for
a time interval equal to half the
length of the LSP so the shell thickness is $v P/2$, where
P is the length of the LSP.  Our models are thus made with two values for the shell
thickness.

The second input parameter we require for \textsc{dusty} is the temperature
$T_{in}$ at the inner edge of the dust shell. We assume $T_{in}$ is the dust condensation temperature.
The dust condensation temperature is commonly assumed to lie in the range of
800--1500 K (e.g. \citealt{van05}; \citealt{cas13}) and in our models we consider
$T_{in}$ values of 800 and 1500 K.   We note that $T_{in}$ determines the inner radius
of the dust shell.  

The third input parameter for \textsc{dusty} is the grain type.
The assumed chemical composition of the grains is different for 
models of oxygen-rich stars and carbon stars.
\citet{van05} used \textsc{dusty} to empirically determine 
possible grain types for oxygen-rich and carbon
AGB stars in the LMC by requiring an acceptable fit to the
SED of the observed stars. We use the grain chemical composition that
they adopted.  For carbon stars, we use dust grains of
50$\%$ amorphous carbon from \citet{han88}, 40$\%$ graphite from
\citet{dra84} and 10 $\%$ SiC from \citet{peg88} and for oxygen-rich
stars, we use the astronomical silicates of \citet{dra84}.

The stellar luminosity $L$ and the effective temperature $T_{\rm eff}$ of our LSP stars
are also required as input to \textsc{dusty}.  These were computed from the
$J$ and $K_s$ magnitudes at maximum light, when circumstellar extinction
is a minimum.  The $T_{\rm eff}$
values for the observed stars were estimated from the $J$-$K_{s}$ color at
maximum light using the formula given in \citet{bes83}.   We also used
$J$-$K_{s}$ to calculate a blackbody temperature $T_{\rm BB}$ for the
central star at maximum.  Luminosities were computed using a bolometric
correction $BC_{K}$ to the $K_s$ magnitude.
For oxygen-rich stars, we adopted the formula
$BC_{K}=0.60+2.56(J-K)-0.67(J-K)^{2}$ given by \citet{ker10} and 
for carbon stars we adopt the
formula $BC_{K}=1.70+1.35(J-K)-0.30(J-K)^{2}$ given by \citet{bes84}.
To estimate of absolute bolometric magnitude, we used the distance 
modulus to the SMC of 18.93 given by \citet{kel06}.  
The derived values for $L$, $T_{\rm BB}$ and $T_{\rm eff}$ are listed in
Table~\ref{tab01}.  When using \textsc{dusty},
we assumed that the central star emits a
blackbody spectrum of temperature $T_{\rm BB}$.  We are only interested in {\em changes} in 
magnitudes and colours, rather than the absolute values of the colours
and magnitudes, and the blackbody assumption will have only an insignificant
effect on these changes.

The final input parameter to \textsc{dusty} is the optical depth $\tau_V$
in the $V$ band.  The output from \textsc{dusty} is the SED of the central
star extincted by a shell of visible optical depth $\tau_V$.
Unless there is significant emission from the dust
shell in each band under consideration, for a given value of $\tau_V$
the extinction of the central star in each band is essentially 
independent of the details of the shell radius or thickness.

\subsection{Comparison of the dust model with observations}
\label{sec:comparison_dust}
\subsubsection{The variation of $I$ with $I-J$}
\label{sec:dust_IJ}
\begin{figure*}
\begin{tabular}{ccc}
\begin{minipage}{0.33\hsize}
\includegraphics[width=1\textwidth]{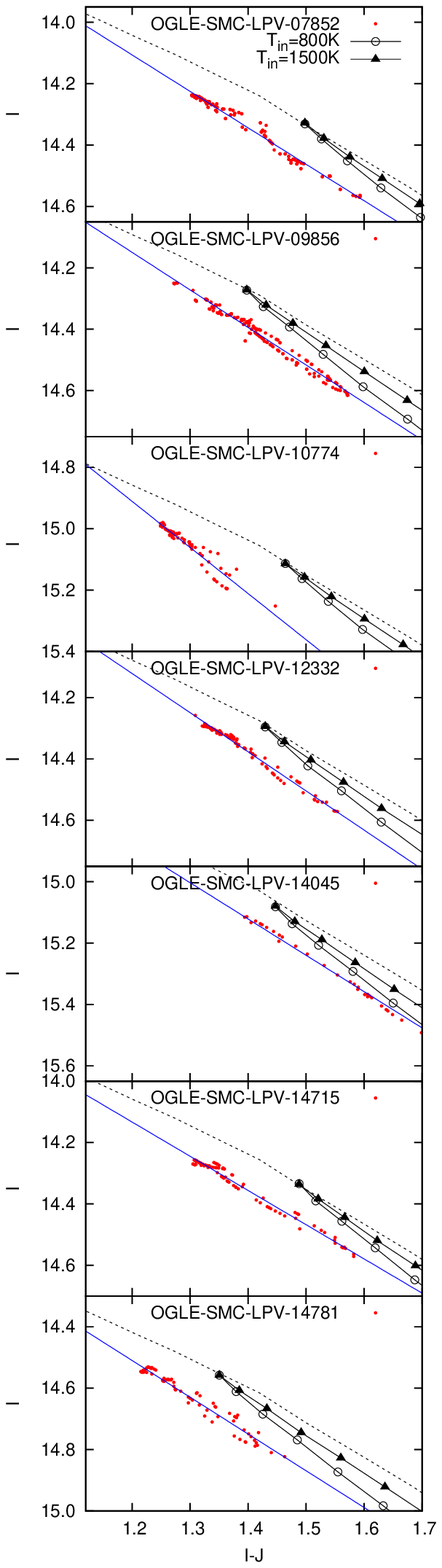}
\end{minipage}
\begin{minipage}{0.33\hsize}
\includegraphics[width=1\textwidth]{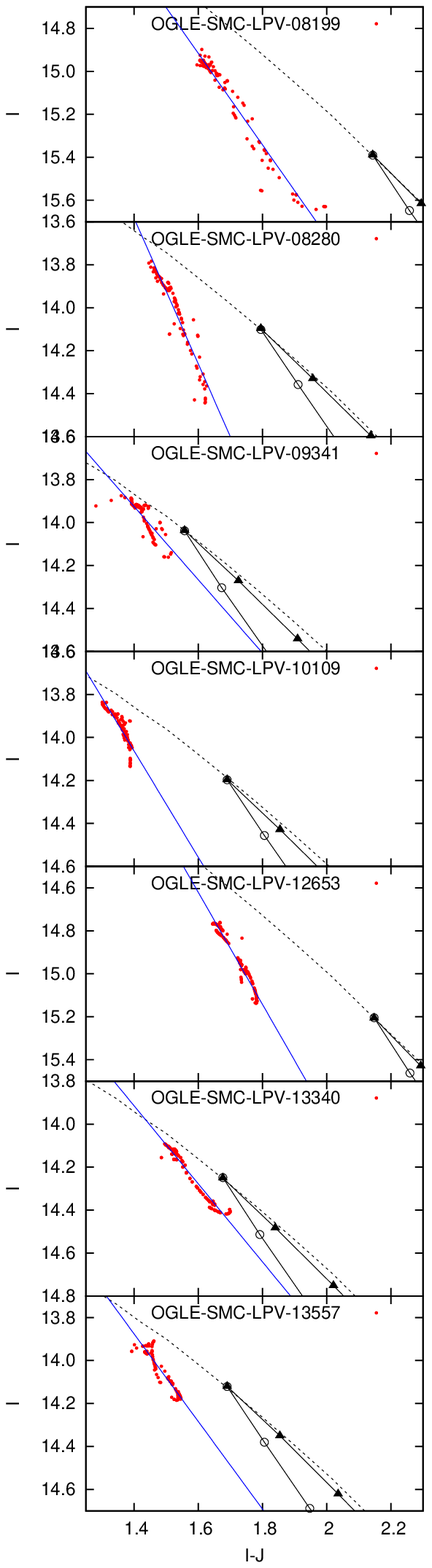}
\end{minipage}
\begin{minipage}{0.33\hsize}
\includegraphics[width=1\textwidth]{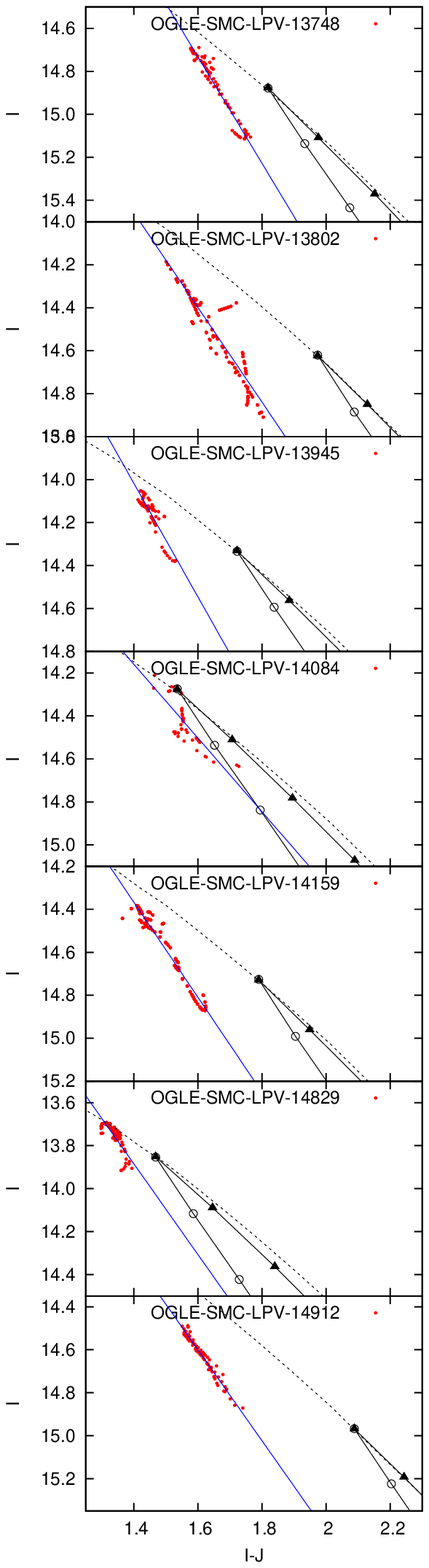}
\end{minipage}
\end{tabular}
\caption{
Comparisons of observations with models of expansion velocity 15 km s$^{-1}$. The
7 panels of the left column represent the $I$ and $I-J$ variations for the
oxygen-rich stars and the 14 panels of the middle and right columns
are for the carbon stars.  The red points show the LSP variations
obtained from Fourier fits to the observed light curves. 
The blue lines show least squares fits to these points (the line is omitted
if the relative error in the slope is more than 0.2).
The black continuous lines with open circles show the magnitude and colour 
variations for models with an
inner dust shell temperature of 800\,K while the triangles are
for models with an inner dust
shell temperature of 1500\,K.  The top end of each curve corresponds to shell optical depth
$\tau_{\rm V} = 0$ and the points are spaced at intervals of 0.5 in $\tau_{\rm V} = 0$.
The black dotted lines show the locus of un-extincted
blackbodies of different temperatures and the constant luminosity
given in Table~\ref{tab01} for each star.  Note that the scale of the vertical axis is the same for all 
oxygen-rich stars and for all carbon stars.
}
\label{I-IJ_15}
\end{figure*}

\begin{figure*}
\begin{tabular}{ccc}
\begin{minipage}{0.33\hsize}
\includegraphics[width=1\textwidth]{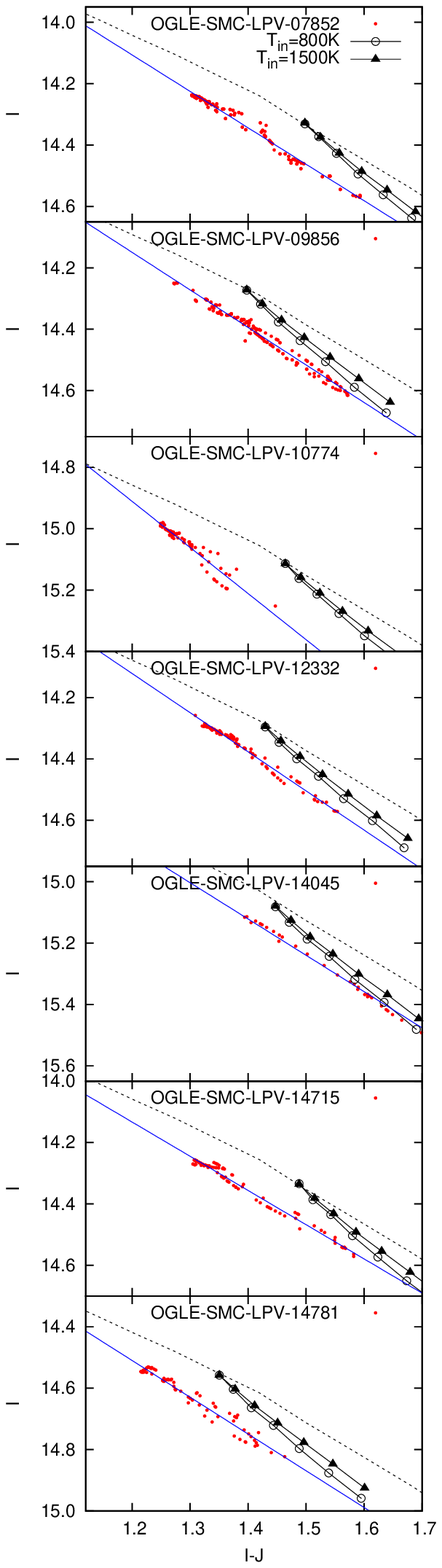}
\end{minipage}
\begin{minipage}{0.33\hsize}
\includegraphics[width=1\textwidth]{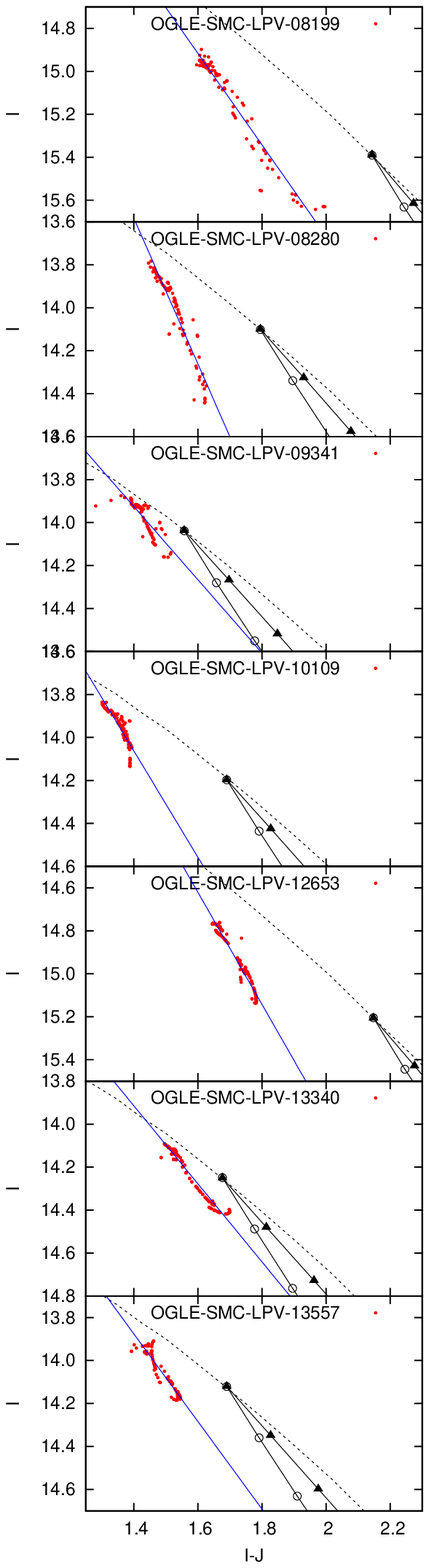}
\end{minipage}
\begin{minipage}{0.33\hsize}
\includegraphics[width=1\textwidth]{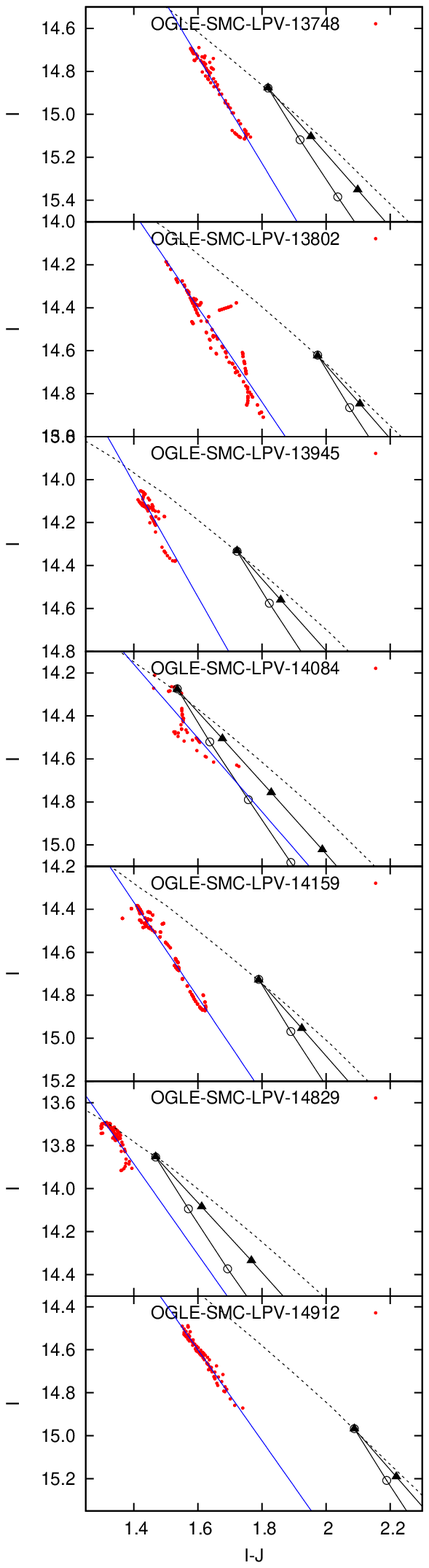}
\end{minipage}
\end{tabular}
\caption{
The same as Figure~\ref{I-IJ_15} but for models
with an expansion velocity of 220 km s$^{-1}$.
}
\label{I-IJ_220}
\end{figure*}

In general, the $I$ light curves are the best sampled of the light
curves in the different filters, the $J$, $H$ and $K$ light curves are
the next best sampled and the $V$ light curves are mostly only sparsely
sampled.  Thus when comparing observations against models in 
magnitude-colour plots, we use $I$ as the magnitude and a colour
involving $I$ and another magnitude.  The other magnitude should be
one of $J$, $H$ or $K$ because of the higher frequency of observation
compared to $V$.  Ideally, $I-K$ would be best because of the
large wavelength range spanned by this colour, but as we will show below, the $K$ band
in O-rich stars is contaminated by variable H$_{\rm 2}$O absorption, as is the $H$ band,
so we use $I-J$ as the colour in our first comparison of observations and
models.

In Figures\,\ref{I-IJ_15} and \ref{I-IJ_220} we show the $I$ magnitude
plotted against the $I-J$ color for the observed samples of oxygen-rich 
and carbon stars.  Figure\,\ref{I-IJ_15} displays models with 
a shell expansion velocity of 15 km s$^{-1}$ (thin shell) while
Figure\,\ref{I-IJ_220} displays models with
a shell expansion velocity of 220 km s$^{-1}$ (thick shell).  
Note that it is not
the absolute values of points in the diagrams that is important, it is 
the slope of the line representing the variation
of $I$ with $I-J$ that matters.  If the LSP variations are caused by variable amounts 
of circumstellar dust alone, then the slope will depend almost entirely on 
the dust properties and not on the spectrum of the central star.
The central star, which in these models we assume to emit as a blackbody
with the $L$ and $T_{\rm BB}$ given in Table~\ref{tab01},
determines the position of the model in Figures\,\ref{I-IJ_15} and \ref{I-IJ_220}
when there is no circumstellar extinction.

Firstly, we make some comments on the models.  In both
Figure\,\ref{I-IJ_15} and Figure\,\ref{I-IJ_220}, the two model lines
differ slightly due to re-emission from the dust shell at $J$ and to a
lesser extent $I$.  Comparing Figure\,\ref{I-IJ_15} with
Figure\,\ref{I-IJ_220}, corresponding to models with different dust
shell thickness, we see that there is a slightly faster increase in $I-J$ with $I$
for the thinner shell of Figure\,\ref{I-IJ_15}, especially for the
hotter inner dust shell temperature of 1500\,K.  The reason for this is that
for the thinner dust shell there is more emission at $J$.  Also, 
for a given optical depth $\tau_V$ to the centre of
the disk of the observed star, the total extinction averaged over the
stellar disk is slightly greater for the thinner shell in
Figure\,\ref{I-IJ_15} than for the thicker shell in Figure\,\ref{I-IJ_220}
(the lines in Figure\,\ref{I-IJ_15} are longer than in
Figure\,\ref{I-IJ_220}).  This is because the extinction near the limb of
the disk is smaller relative to the extinction at the centre of the
disk for a thicker shell.  

The observed variation of $I$ with $I-J$ associated with the LSP 
(red points) is
generally consistent with the slope of one of the model lines in
Figures\,\ref{I-IJ_15} or \ref{I-IJ_220}.  This indicates that the
variation of $I$ and $J$ due to an LSP could be explained by 
variable dust absorption alone.

\subsubsection{The variation of $I$ with $J-K_{s}$}\label{sec:dust_JKs}

\begin{figure*}
\begin{tabular}{ccc}
\begin{minipage}{0.33\hsize}
\includegraphics[width=1\textwidth]{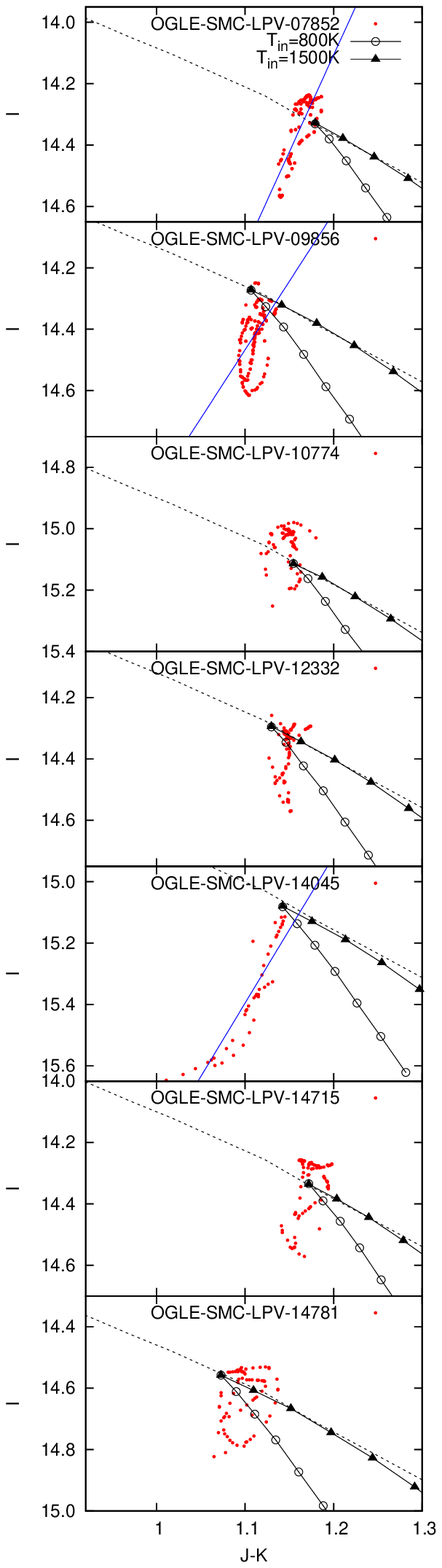}
\end{minipage}
\begin{minipage}{0.33\hsize}
\includegraphics[width=1\textwidth]{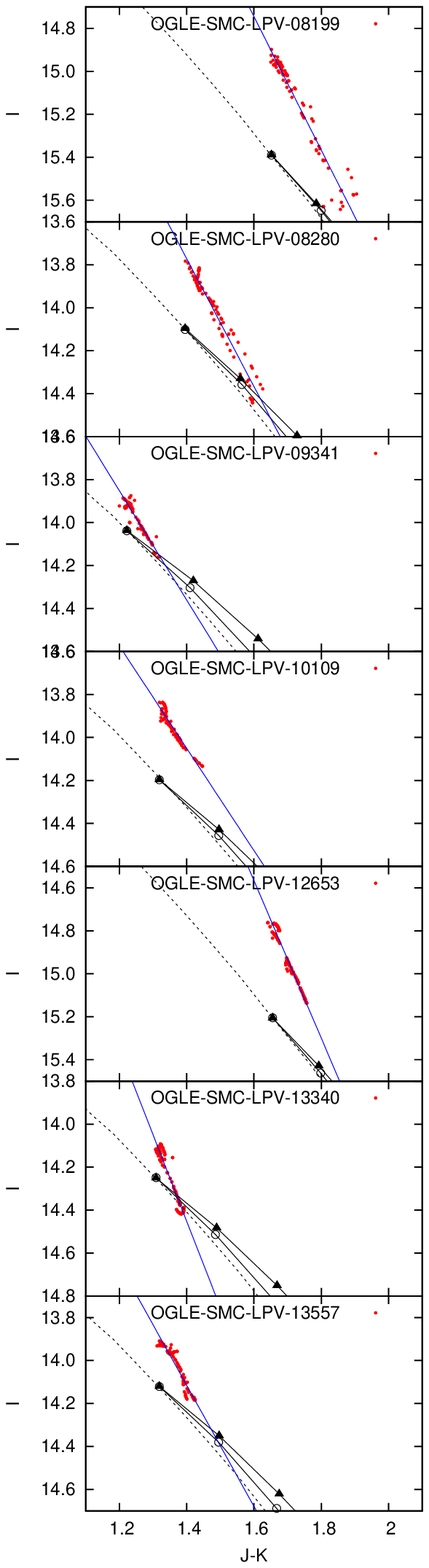}
\end{minipage}
\begin{minipage}{0.33\hsize}
\includegraphics[width=1\textwidth]{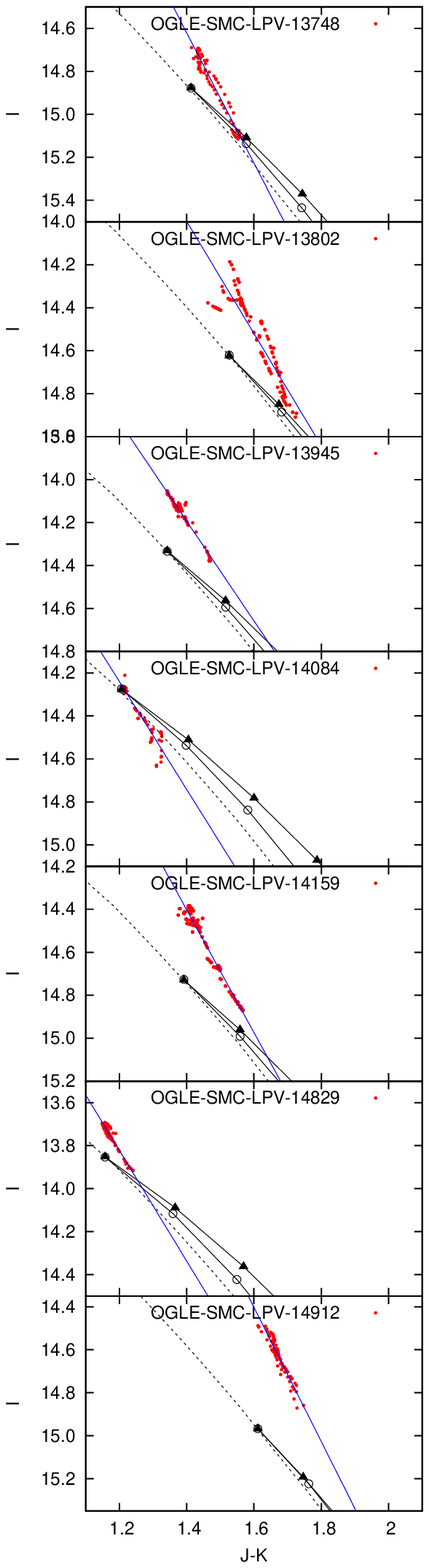}
\end{minipage}
\end{tabular}
\caption{The same as Figure~\ref{I-IJ_15} but using the $J-Ks$ colour
rather than $I-J$.}
\label{I-JK_15}
\end{figure*}

\begin{figure*}
\begin{tabular}{ccc}
\begin{minipage}{0.33\hsize}
\includegraphics[width=1\textwidth]{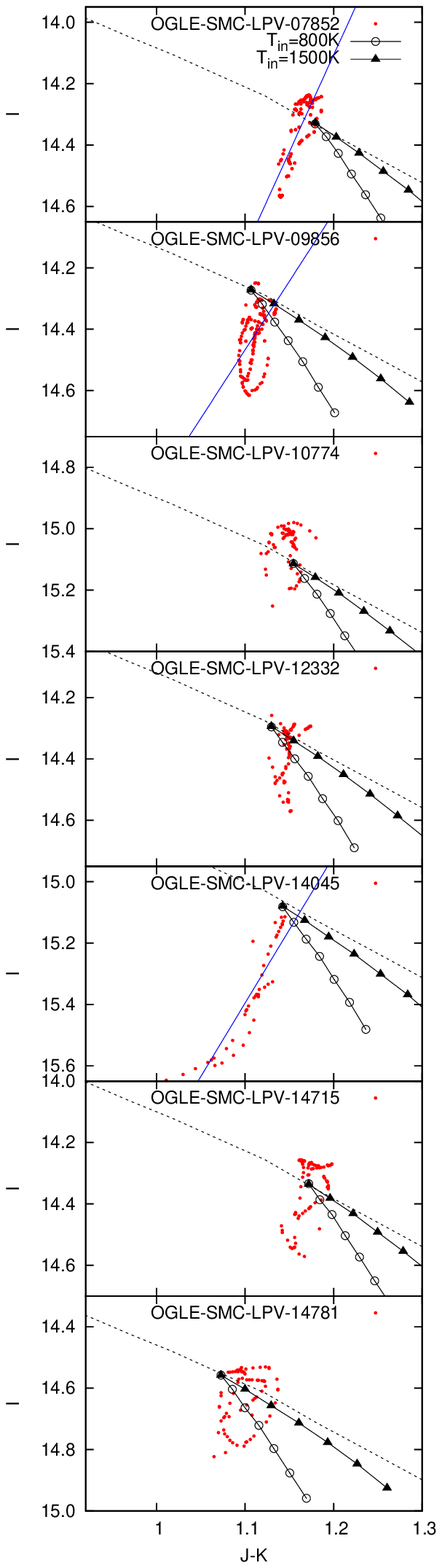}
\end{minipage}
\begin{minipage}{0.33\hsize}
\includegraphics[width=1\textwidth]{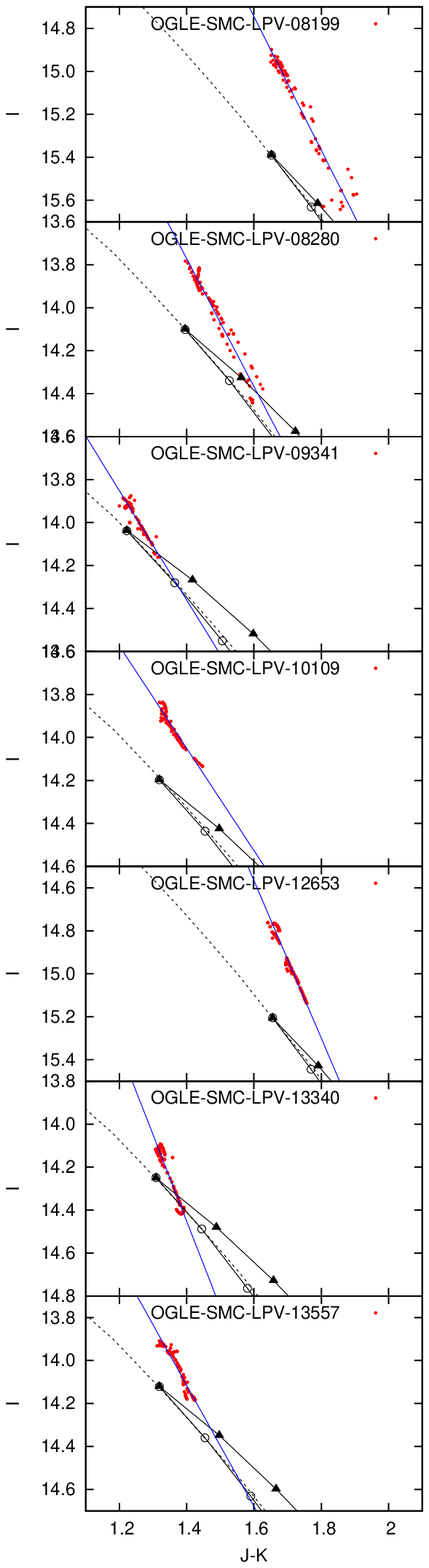}
\end{minipage}
\begin{minipage}{0.33\hsize}
\includegraphics[width=1\textwidth]{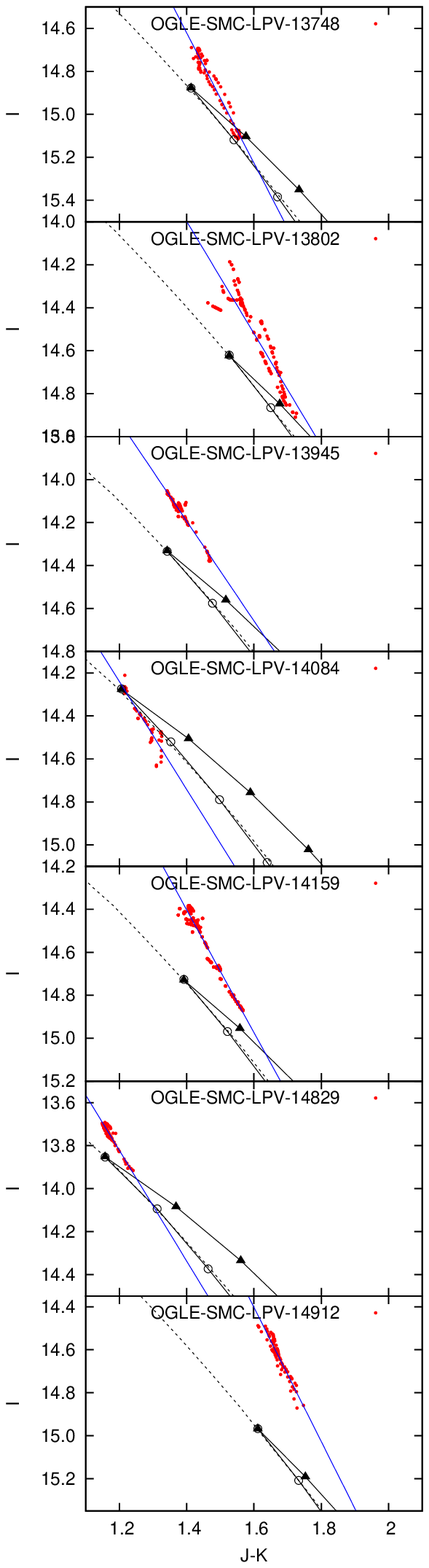}
\end{minipage}
\end{tabular}
\caption{The same as Figure~\ref{I-IJ_220} but using the $J-Ks$ colour
rather than $I-J$.}
\label{I-JK_220}
\end{figure*}

Figures\,\ref{I-JK_15} and \ref{I-JK_220} show the variation of $I$
with $J-K_{s}$.  As one can easily see, the observational variations 
due to the LSPs are inconsistent with the model variations in all cases.  Most
importantly, for the O-rich stars, the observed $J-K_s$ colour tends to get 
{\em bluer as the star gets fainter in $I$}.  There is no way that
such variations can be the result of variable amounts of dust extinction
only.  

The most probable explanation for the variation of $J-K_s$ in O-rich
LSP stars is strong H$_{\rm 2}$O absorption in $K_{s}$ band.  Significant
broad H$_{\rm 2}$O band absorption is found between about 1.4$\mu$m and
1.9$\mu$m on the edges of the $H$ and $K_{s}$ filter bandpasses.  In
non-variable stars, this absorption is seen from about M6
I\hspace{-.1em}I\hspace{-.1em}I to later spectral type in M giants
(e.g. \citealt{tsu00}; \citealt{ray09}).  Strong water absorption in
the $H$ and $K$ bands is also commonly seen in large amplitude pulsating red giants
(Mira variables) \citep[e.g.][]{jm70}.  The reason for this
absorption is the existence of a relatively dense shell with a temperature of
about 1800\,K elevated above the photosphere by shock waves associated
with the pulsation \citep{isw11}.  Using the models atmosphere results
of \citet{hou00}, we find that the $V-I$ colours of the stars we are
examining here yield $T_{\rm eff} \sim $3400--3600\,K, corresponding
to spectral types M3--M5.  Due to these relatively early M spectral types, 
the results noted above mean that strong water
absorption in the $K_s$ band can not occur in a hydrostatic atmosphere.  
The elevation of a shell of mass
above the photosphere is required to produce strong water
absorption, as in the case of Mira variables.  We suggest
that this shell elevation, which may also lead to dust formation,
occurs at the beginning of the luminosity declines associated with the
LSPs, leading to the formation of increasing $H_{\rm 2}$O absorption in
the $K_s$ band and blueing of the $J-K_s$ colour.

This result is a second piece of observational evidence for the existence of
a disturbance above the photosphere caused by the LSP. \citet{woo04}
found variable H$\alpha$ absorption in stars with LSPs which they
attribute to a chromosphere of variable strength, with the strongest
H$\alpha$ absorption near maximum light.  Combined with the results
above for the H$_{\rm 2}$O absorption layer, we have a picture whereby
matter is ejected above the photosphere beginning near the luminosity
decline of the LSP forming a relatively dense shell of temperature
$\sim$1800\,K.  As the LSP cycle progresses, this shell appears to
be slowly heated by non-thermal processes to $\sim$8000\,K giving
rise to a chromosphere which is then displaced by the next ejected shell
containing H$_{\rm 2}$O.

Turning to the carbon stars, we see that they do get redder as they get fainter in $I$ but in all
cases they clearly do not redden as much as predicted by the models.  
To bring the observed variation in $J-K_s$ with decline in $I$ into 
agreement with the models, while keeping the variation of $I-J$ with $I$
in agreement with the models, would require a strange dust opacity with a large bump in
the $K_s$ band.  We do not know of any dust composition that could produce
such a bump (see Section~\ref{sec:dust_composition} for a discussion of the effect
of different grain opacities).

We note that ejected clouds covering only part of the stellar surface
rather than the complete surface as for spherical shells can not
explain the observed variation 
in $J-K_s$ with declining $I$.  For all but very
large optical depths, the $J-K_s$ colour in a cloud model increases
slightly faster with decline in $I$ than for a spherical shell model,
whereas the observations require a slower increase in $J-K_s$ colour.
Also, given that the variation of the observed $I-J$ colour with
decline in $I$ agrees with the spherical shell models, a cloud model
will not agree with the observations (the $V-I$ colour in the cloud
model increases more slowly with decline in $I$ than in the spherical
shell model).

In summary, our conclusion from examining the variations of LSP stars in the
($I$,$J-K$) plane is that, for both O-rich and C-rich stars, a model
involving only variable dust extinction is not viable as an explanation
for LSP behaviour.

\subsubsection{The variation of $I$ with $V-I$}

\begin{figure*}
\begin{tabular}{ccc}
\begin{minipage}{0.33\hsize}
\includegraphics[width=1\textwidth]{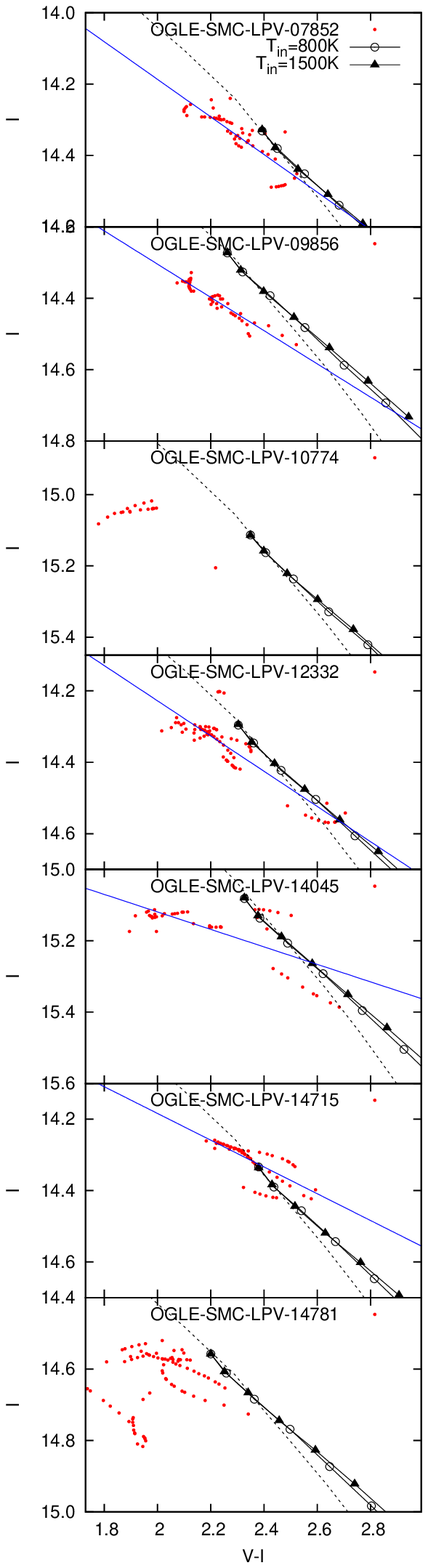}
\end{minipage}
\begin{minipage}{0.33\hsize}
\includegraphics[width=1\textwidth]{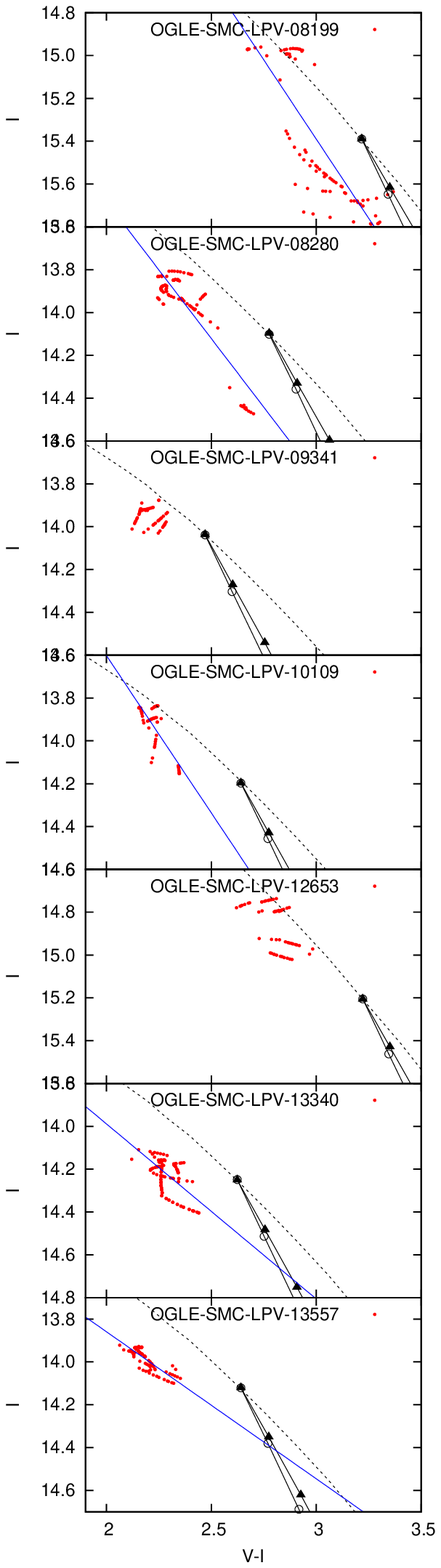}
\end{minipage}
\begin{minipage}{0.33\hsize}
\includegraphics[width=1\textwidth]{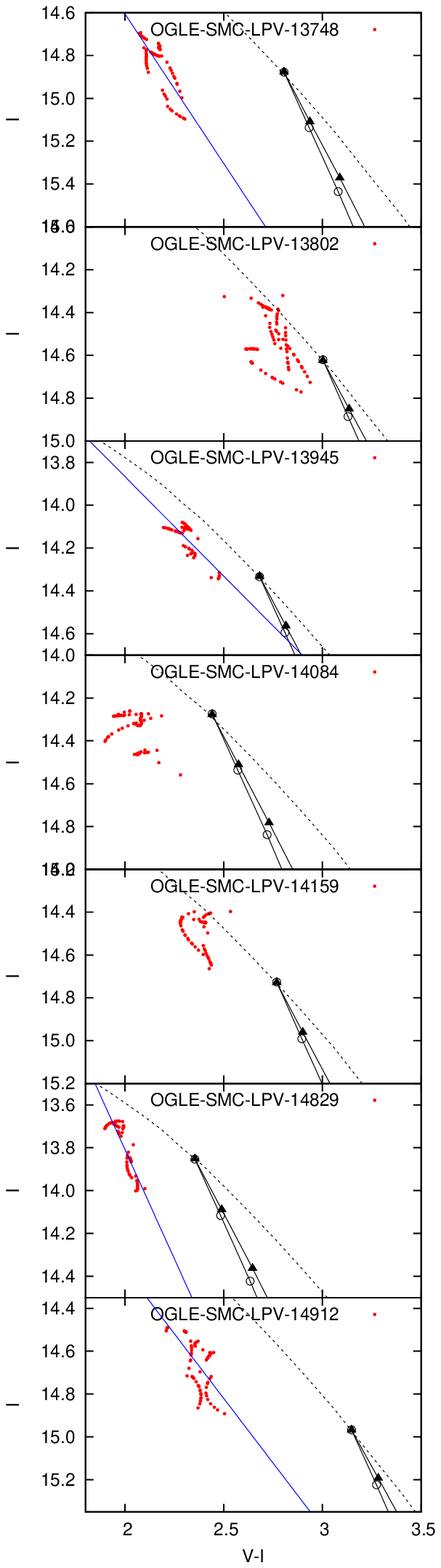}
\end{minipage}
\end{tabular}
\caption{
The same as Figure~\ref{I-IJ_15} but using the $V-I$ colour
rather than $I-J$.}
\label{I-VI_15}
\end{figure*}

\begin{figure*}
\begin{tabular}{ccc}
\begin{minipage}{0.33\hsize}
\includegraphics[width=1\textwidth]{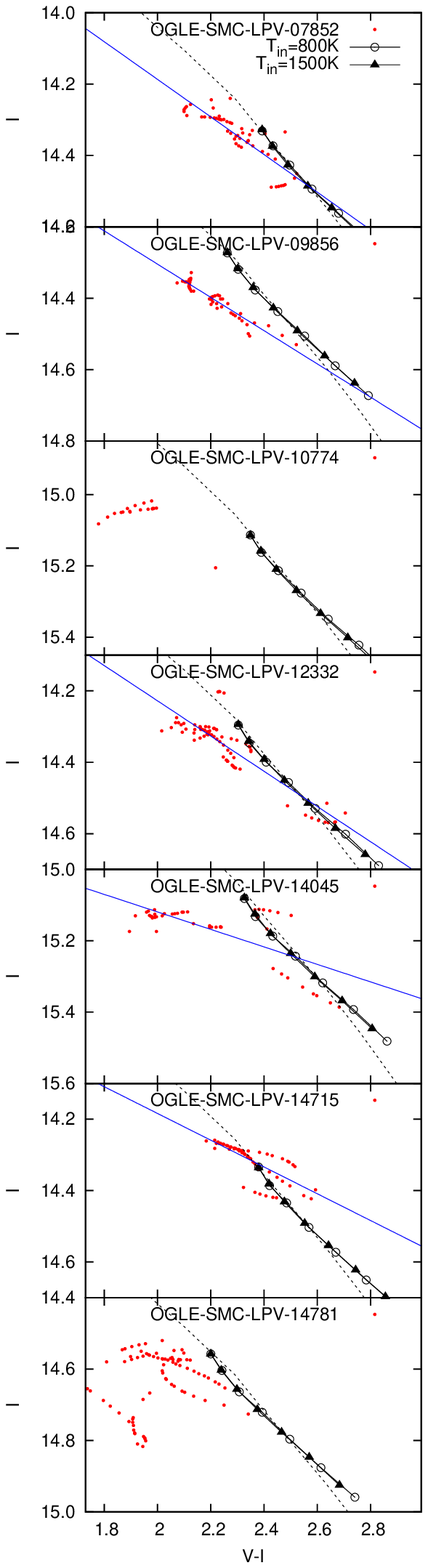}
\end{minipage}
\begin{minipage}{0.33\hsize}
\includegraphics[width=1\textwidth]{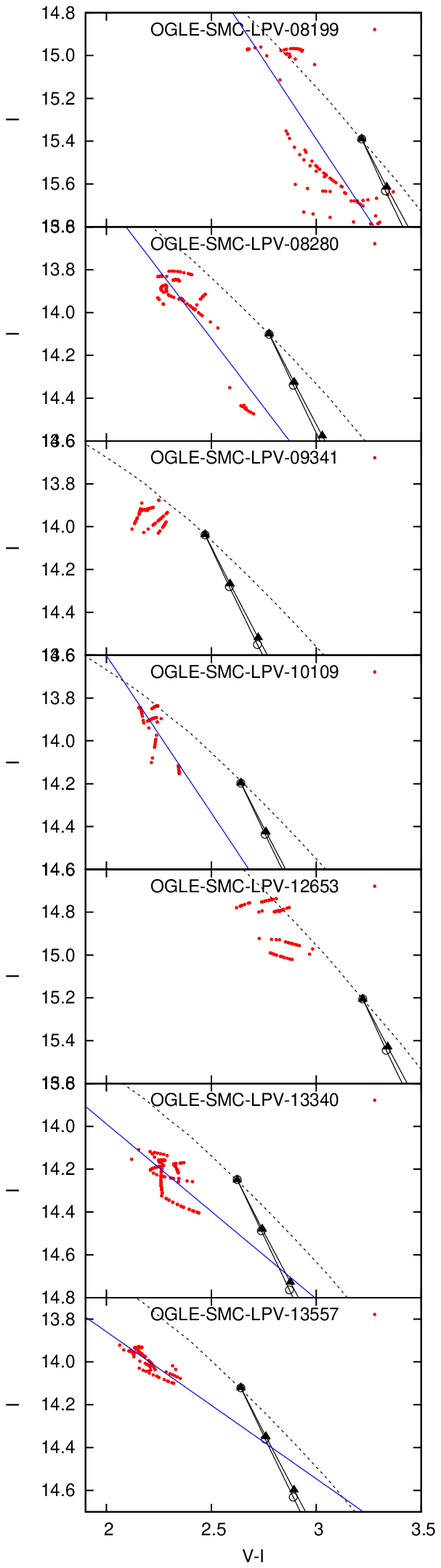}
\end{minipage}
\begin{minipage}{0.33\hsize}
\includegraphics[width=1\textwidth]{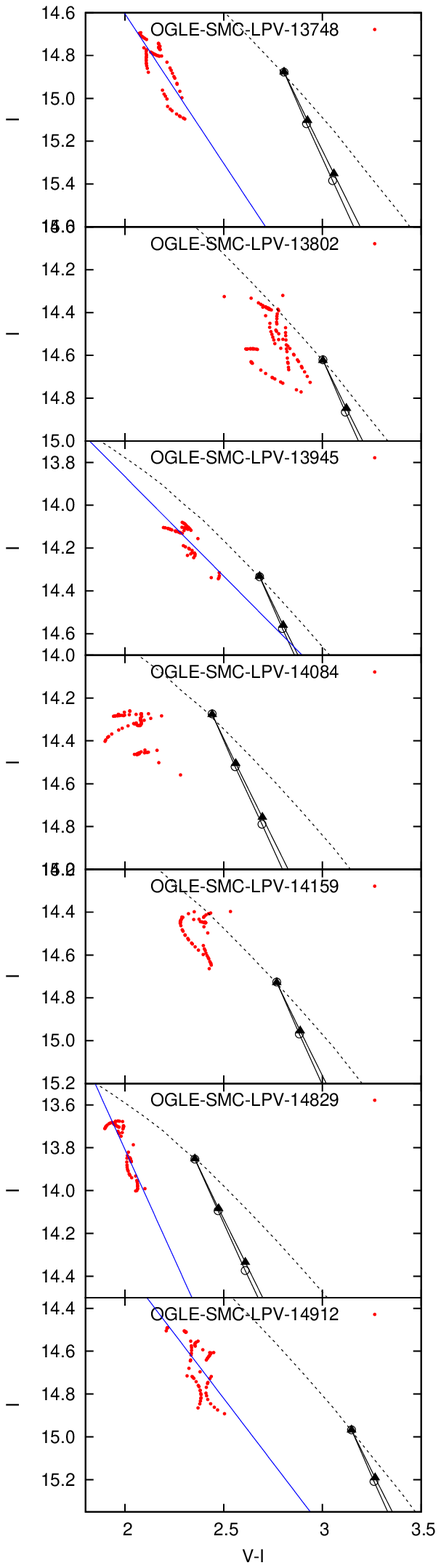}
\end{minipage}
\end{tabular}
\caption{
The same as Figure~\ref{I-IJ_220} but using the $V-I$ colour
rather than $I-J$.}
\label{I-VI_220}
\end{figure*}

Although the $V$ observations are not as frequent as observations
in other bands, we show the variation of $I$ with $V-I$ for completeness
in Figures~\ref{I-VI_15} and \ref{I-VI_220}.  For both the $V$ and $I$ bands,
re-emission of light absorbed by the shell is weak so the model
lines corresponding to the two inner shell temperatures of 800\,K and 1500\,K
are close together in both figures.

For most O-rich stars, the observed $V-I$ tend to increase slightly more rapidly
with declining $I$ than in the models.  For most C-rich stars,
the observed  and model variations have consistent slopes although for
OGLE-SMC-LPV-13557 and OGLE-SMC-LPV-13945 the slopes are different.
Overall, the agreement between the observations and dust model predictions
is not good in the ($I$,$V-I$) plane but the results are not definitive.

\subsubsection{Testing different dust grains}
\label{sec:dust_composition}

Different dust grains have different absorption and scattering
coefficients.  Thus we tried changing the dust grain composition
to see if this would give agreement between the light variations
in the dust shell model and the
observed light variations, especially in the $I$ and $J-K_{s}$ plane for
the carbon stars. We examined four dust grain types available in \textsc{dusty} (see
Table~\ref{tab02}) for oxygen-rich stars and for carbon stars, including
the grain types we used in Section~\ref{sec:model}.  The thin dust shell
model with an expansion velocity of 15 km s$^{-1}$ was
adopted.

\begin{figure*}
\begin{tabular}{ccc}
\begin{minipage}{0.33\hsize}
\includegraphics[width=1\textwidth]{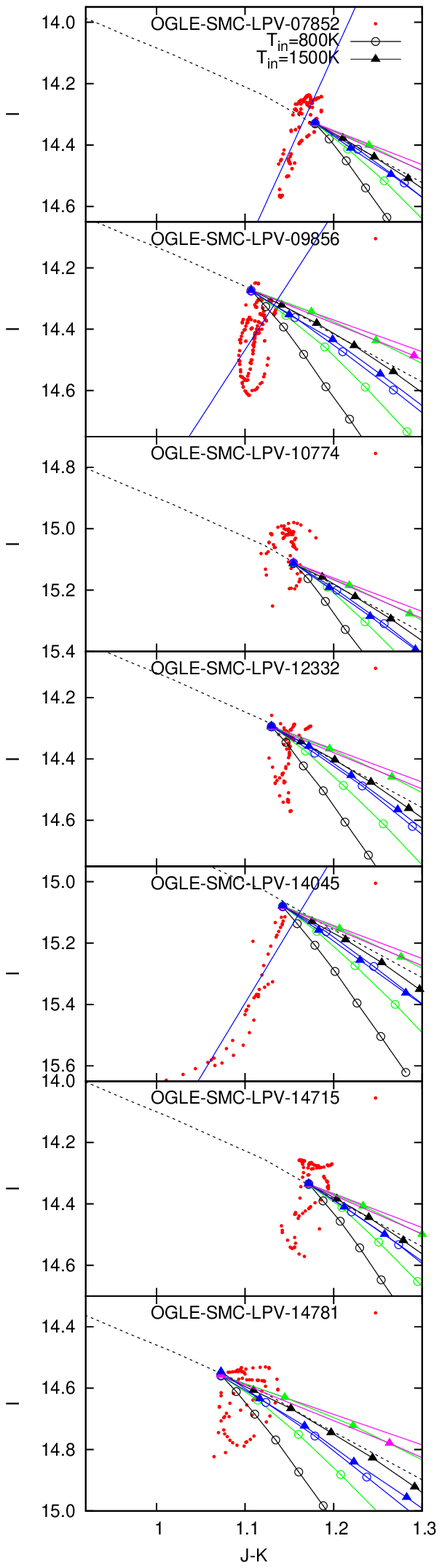}
\end{minipage}
\begin{minipage}{0.33\hsize}
\includegraphics[width=1\textwidth]{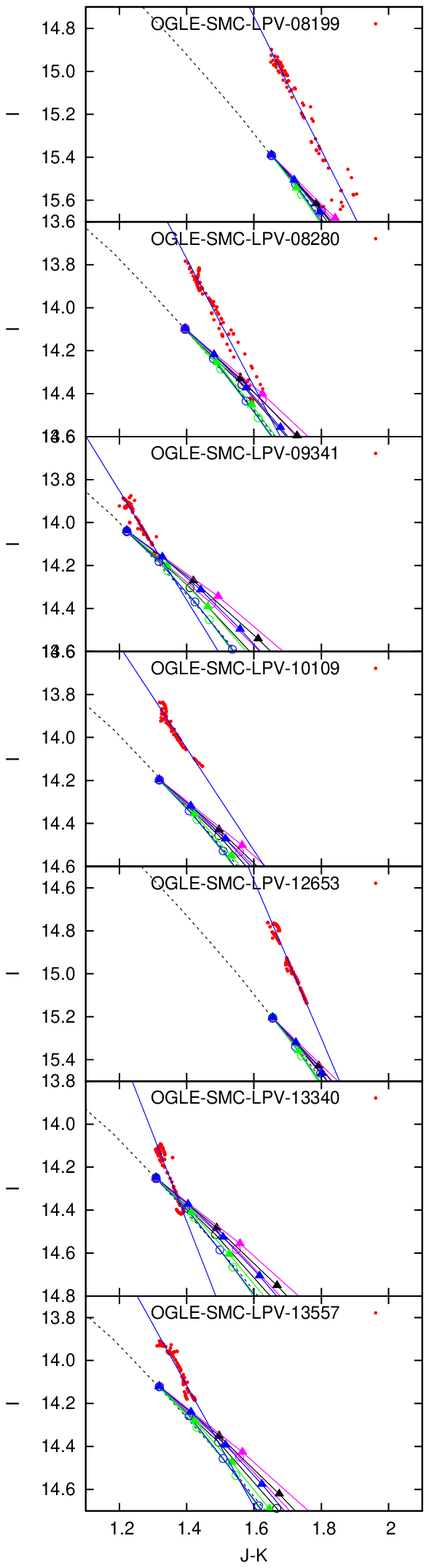}
\end{minipage}
\begin{minipage}{0.33\hsize}
\includegraphics[width=1\textwidth]{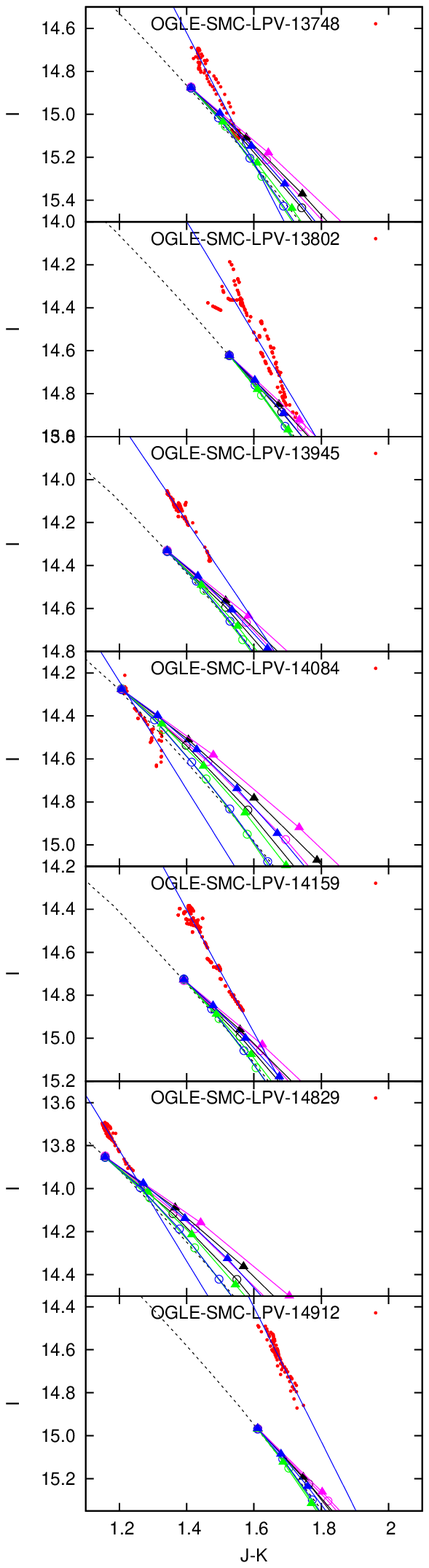}
\end{minipage}
\end{tabular}
\caption{
The same as Figure~\ref{I-JK_15} but
including different dust compositions. For both oxygen-rich 
and carbon stars, the black solid lines with 
open circles and filled triangles correspond to the original
dust compositions (model 0 in Table~\ref{tab02}), while the green, magenta and blue 
colours correspond to models 1, 2 and 3, respectively.
}
\label{I-JK_15_comp}
\end{figure*}

Figure~\ref{I-JK_15_comp} shows comparisons of observed variations 
in the $I$ and $J-K_{s}$ diagram with dust shell models 
for the different grain types. The model slopes for oxygen-rich stars can be changed
appreciably by different dust grains but none of the models
reproduce the observed slopes.  In particular, none of the models can
explain the observation that the $J-K_{s}$
colour get bluer as the star gets fainter in $I$. 
Turning to the carbon stars in Figure~\ref{I-JK_15_comp}, 
we see that changing the grain type in the models changes the slope of the 
variations but again none of the adopted grain types can
consistently reproduce the observed slope.  These results strengthen the finding
of Section~\ref{sec:dust_JKs} that dust shell models can not by themselves
explain the broadband light and colour variations in LSP stars.

\begin{table}
\begin{center}
\begin{minipage}{80mm}
\caption{Dust composition}
\begin{tabular}{ccc}
\hline
\hline
Sp. type& model name & composition \\
\hline
O rich star & model 0 & 100$\%$ $^1$astronomical silicate\\
& model 1 & 100$\%$ $^2$warm O-deficient silicate\\
& model 2 & 100$\%$ $^3$FeO\\
& model 3 & 100$\%$ $^4$glassy olivine\\
\hline
C rich star & model 0 & 50$\%$ $^5$amorphous carbon,\\ 
& & 40$\%$ $^1$graphite, 10$\%$ $^6$$\alpha$-SiC\\
& model 1 & 100$\%$ $^1$graphite\\
& model 2 & 100$\%$ $^5$amorphous carbon\\
& model 3 & 100$\%$ $^6$$\alpha$-SiC\\
\hline
\end{tabular}
Note: $^1$\citet{dra84}, $^2$\citet{oss92}, $^3$\citet{hen95}, $^4$\citet{dor95}, $^5$\citet{han88}, $^6$\cite{peg88}
\label{tab02}
\end{minipage}
\end{center}
\end{table}

\subsection{The dark spot model} \label{sec:spot_model}

Another possible explanation of the LSP phenomenon is a 
dark spot on a rotating AGB
star. \citet{sos14} noted that the observed light curves of LSP stars
are similar to those of the known spotted stars. 
Moreover, the $V$ to $I$ amplitude ratios of LSP stars and known spotted stars
are similar and they are different from the ratios seen in the ellipsoidal binaries of sequence E
which lie near the LSP stars in period-luminosity plots.
Here we test a simple rotating dark spot model for LSP stars using the latest
version of the Wilson-Devinney code \citep{wil71} which has
the ability to simulate the light curve of a spotted star in a binary system.
Since our aim is to examine the light and colour variations due to
the appearance of a dark spot on a red giant, our binary systems
consist of a rotating red giant star with a circular isothermal spot and
a very faint orbiting companion which can not be seen.  Hence, 
we obtain the light and colour variations due to only the dark spot on the rotating red giant.
For the red giant
star, we assume a mass of $\sim$1.0$M_{\odot}$ and  
a radius of $\sim$170$R_{\odot}$, 
which gives a surface gravity $\log g$ [cgs] of $-0.02$. 
A metallicity, $\log$[Fe/H]
= $-1.0$ and a gravity-darkening coefficient, $g_{1}$=0.32
\citep{luc67} are used. For the calculation of oxygen-rich stars, we use
the model stellar atmospheres of \citet{van_h03} while for carbon stars,
a blackbody is used. We note that the model atmospheres do not take into
account the effect of the absorption due to H$_{\rm 2}$O molecules. 

The amplitudes of the light variations due to the presence of spots are determined by
the spot sizes and the temperature differences  
between the spots and the normal stellar surface. The
temperatures of spots on AGB stars are not known
well.  For the spot temperature, we follow the work of \citet{sok99}.  They initially
adopted the solar value of $\frac{2}{3}$ for the ratio of spot to normal photospheric temperature.
For an assumed photospheric temperature of 3000\,K, the spot temperature would then be
2000\,K.  \citet{sok99} then considered a range of spot temperatures from 1600\,K (cool spot,) and 2500\,K (warm spot)
corresponding to a ratio of spot to normal photospheric temperature of 0.533 and 0.833, respectively.
To cover a range of spot temperatures, we also make models with these two temperature ratios.
We use these models to simulate each of the stars in Table~\ref{tab01} and we use
the effective temperatures $T_{\rm eff}$ in Table~\ref{tab01} as the temperature of
the normal photospheric regions.

The spot sizes for the two temperature cases are fixed by requiring that
the amplitude in the $I$ band is similar to the observed largest amplitudes,
which are $\sim$0.5 magnitudes for the oxygen-rich stars and
$\sim$0.7 magnitudes for the carbon stars.  Table~\ref{tab03}
gives the parameters of our spot models.  It has been assumed that the
inclination of the model rotation axis is 90 degrees to the line of sight and the 
center of the spot is on the equator of the star.

The spot model also effectively simulates  
light variations caused by the giant
convective cells which were proposed by \citep{sto10} to explain
LSPs in AGB stars.
In this model, the light variations are
caused by the turnover of cooler and hotter 
giant convective cells at the stellar surface.  To simulate this situation,
we examine a model where the rotating dark spot covers a full hemisphere
of the star (hereinafter, the hemisphere model).  To reproduce
the observed largest amplitudes in $I$ band, we use the
temperature ratios of the spot area to the normal area
given by Table~\ref{tab03}.  As in the case of the other
spot models, we use the effective temperatures $T_{\rm eff}$ given
in Table~\ref{tab01} for the temperature of the normal 
photospheric regions in both
oxygen-rich and carbon stars.

\begin{table}
\begin{center}
\begin{minipage}{90mm}
\caption{Spot temperatures and the spot sizes}
\begin{tabular}{cccc}
\hline
\hline
Sp. type& temperature ratio & spot radius($^\circ$) & model \\
\hline
O rich star & 0.533 & 35 & cool\\
& 0.833 &45 & warm \\
& 0.913 & 90 & hemisphere\\
\hline
C rich star & 0.533 & 40 & cool\\
& 0.833 & 50 & warm\\
& 0.888 & 90 & hemisphere\\
\hline
\end{tabular}
 Note: Spot radius is defined as the angle corresponding to the spot radius when
viewed from the stellar center.
\label{tab03}
\end{minipage}
\end{center}
\end{table}
 
\subsection{Comparison of the spot model with observations}
\label{sec:comparison_spot}
\subsubsection{The variation of $I$ with colours}

\begin{figure*}
\begin{tabular}{ccc}
\begin{minipage}{0.33\hsize}
\includegraphics[width=1\textwidth]{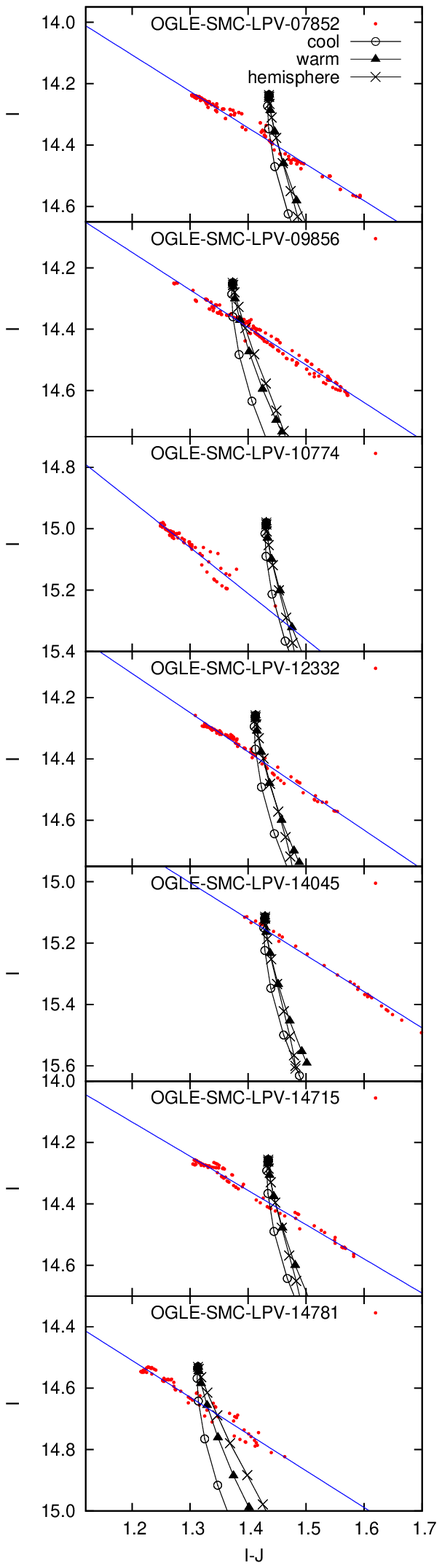}
\end{minipage}
\begin{minipage}{0.33\hsize}
\includegraphics[width=1\textwidth]{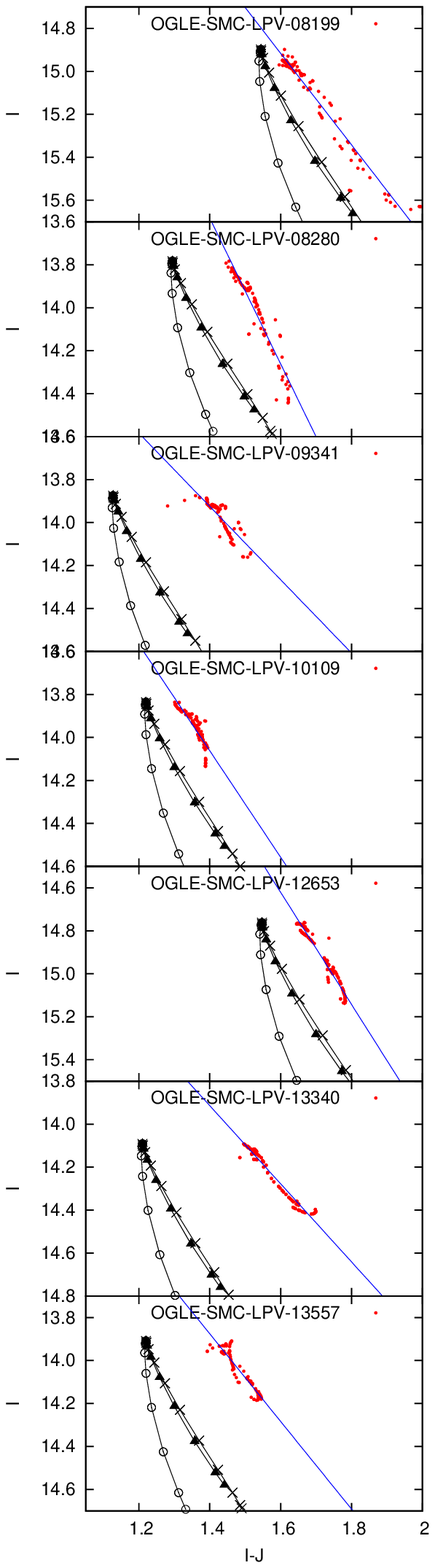}
\end{minipage}
\begin{minipage}{0.33\hsize}
\includegraphics[width=1\textwidth]{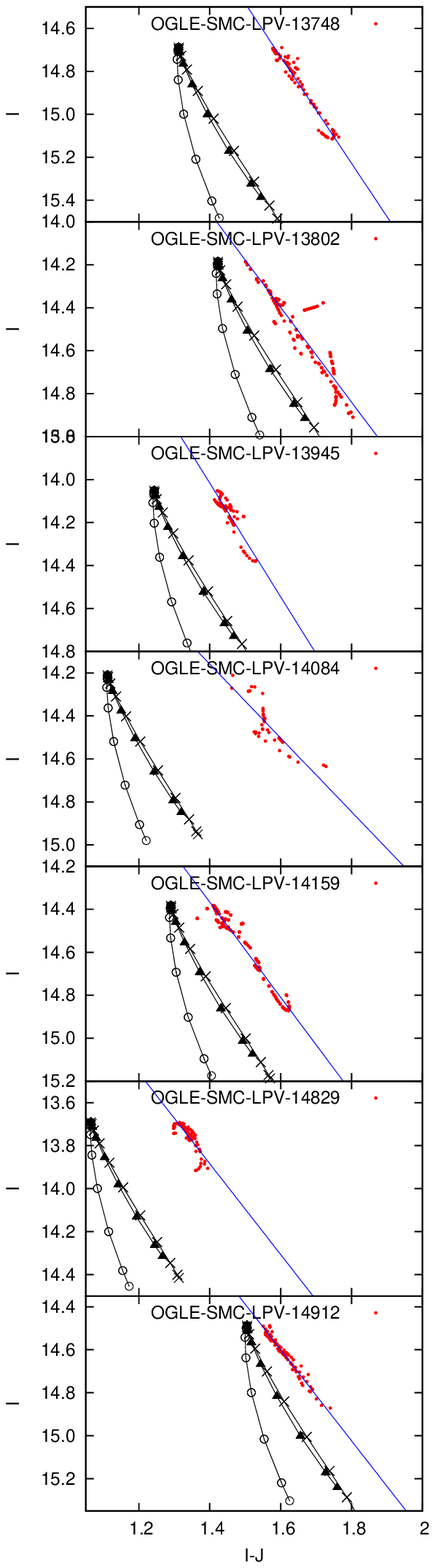}
\end{minipage}
\end{tabular}
\caption{
Comparisons of observations with dark spot models in the ($I$,$I-J$)
plane. The 7 panels of the left column show oxygen-rich stars and the
14 panels of middle and right columns show carbon stars.  Red points
and blue lines represent observations, as in Figure \ref{I-IJ_15}.  The black
lines with open circles and the lines with filled triangles show the
magnitude and colour variations for models with cool and warm spot
temperatures, respectively, while the lines with crosses are for models with a
hemisphere spot.
}
\label{spot_IJ}
\end{figure*}

\begin{figure*}
\begin{tabular}{ccc}
\begin{minipage}{0.33\hsize}
\includegraphics[width=1\textwidth]{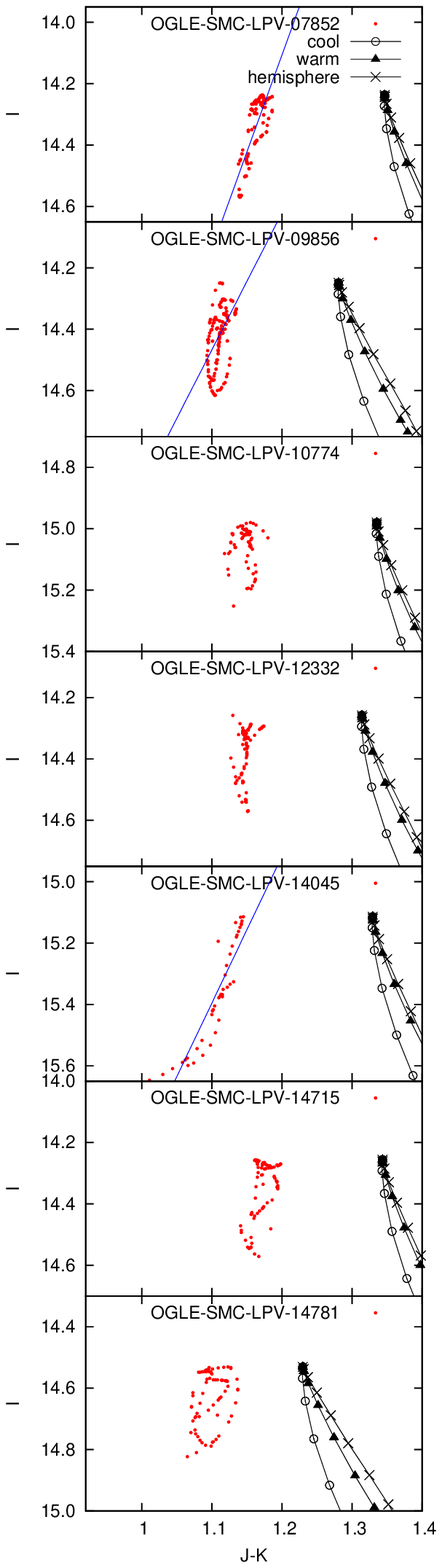}
\end{minipage}
\begin{minipage}{0.33\hsize}
\includegraphics[width=1\textwidth]{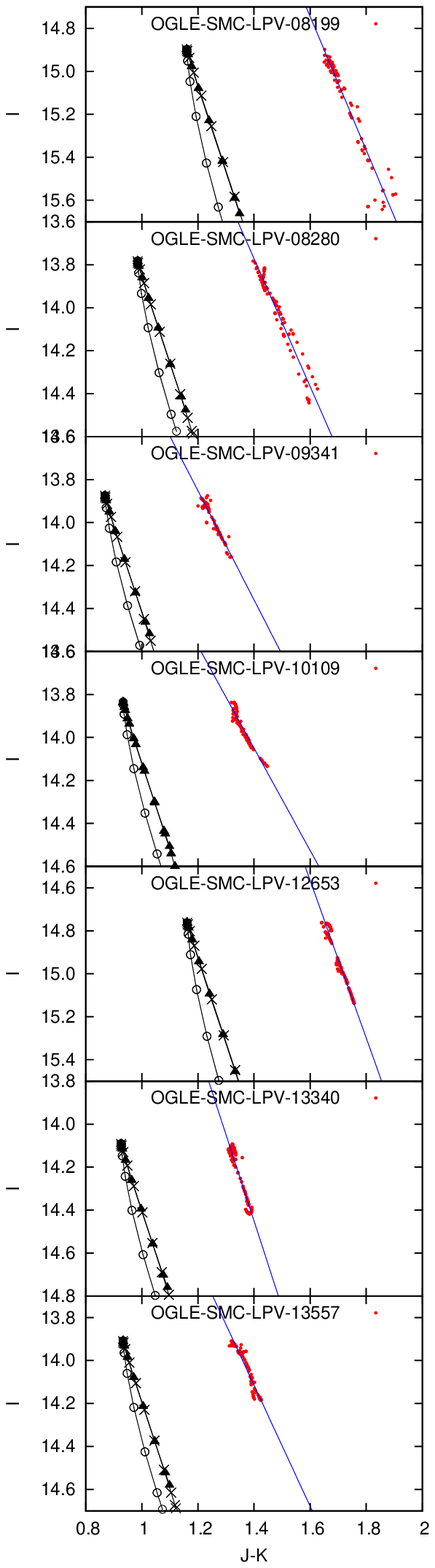}
\end{minipage}
\begin{minipage}{0.33\hsize}
\includegraphics[width=1\textwidth]{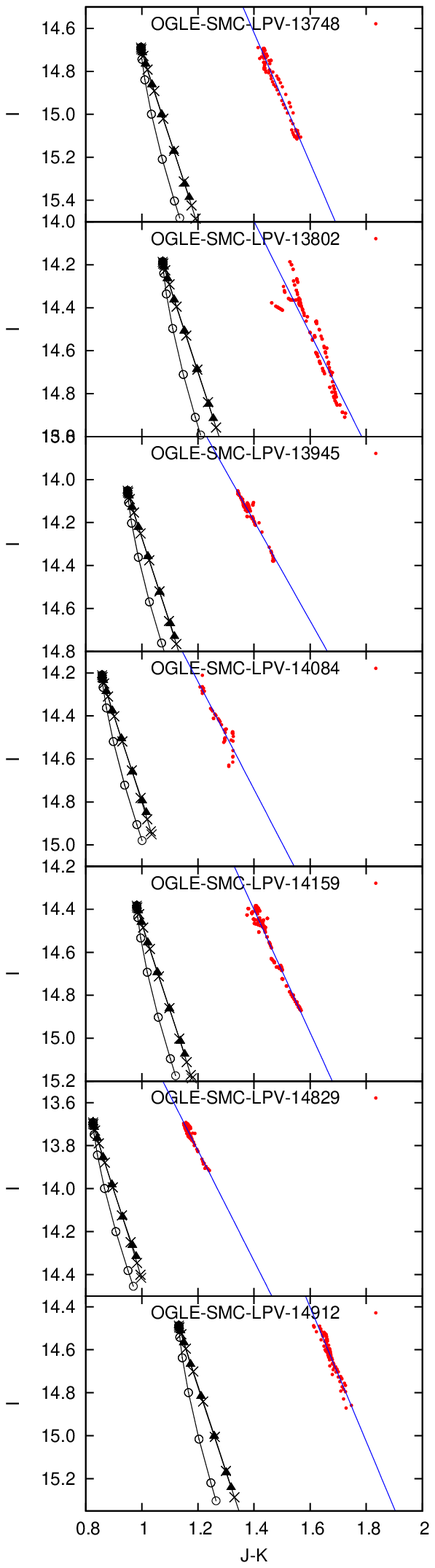}
\end{minipage}
\end{tabular}
\caption{
 The same as Figure \ref{spot_IJ} but using the $J-K_{s}$
  colour rather than $I-J$.
}
\label{spot_JK}
\end{figure*}

\begin{figure*}
\begin{tabular}{ccc}
\begin{minipage}{0.33\hsize}
\includegraphics[width=1\textwidth]{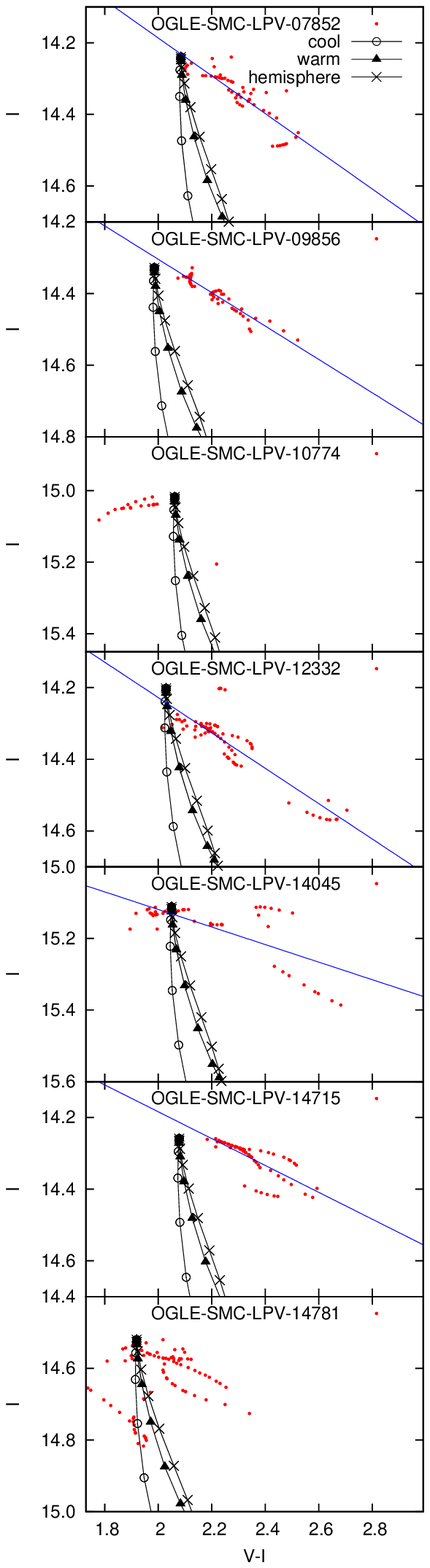}
\end{minipage}
\begin{minipage}{0.33\hsize}
\includegraphics[width=1\textwidth]{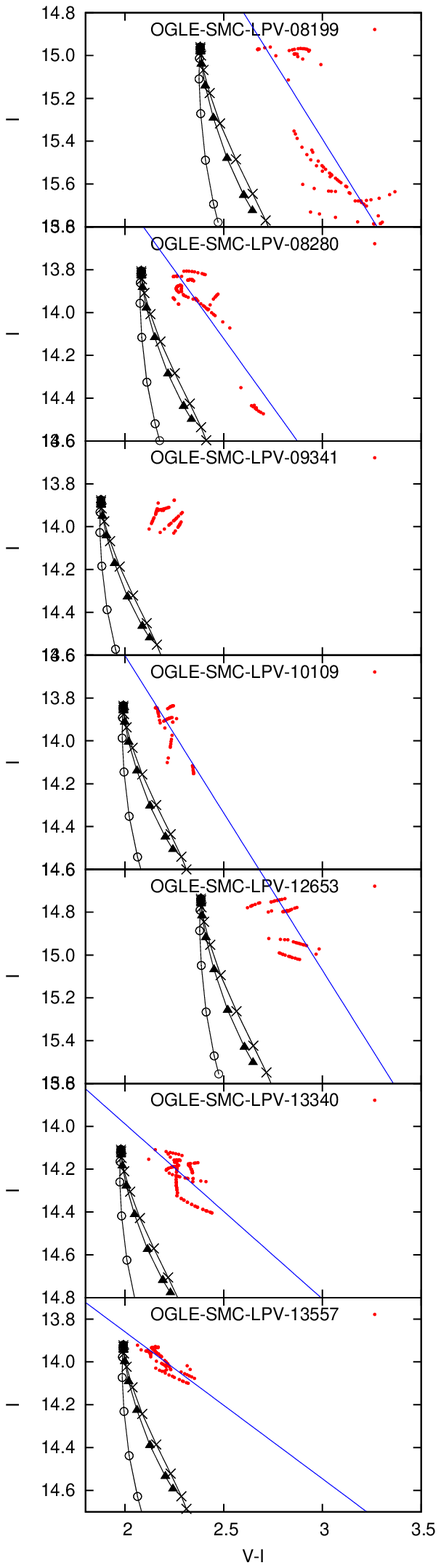}
\end{minipage}
\begin{minipage}{0.33\hsize}
\includegraphics[width=1\textwidth]{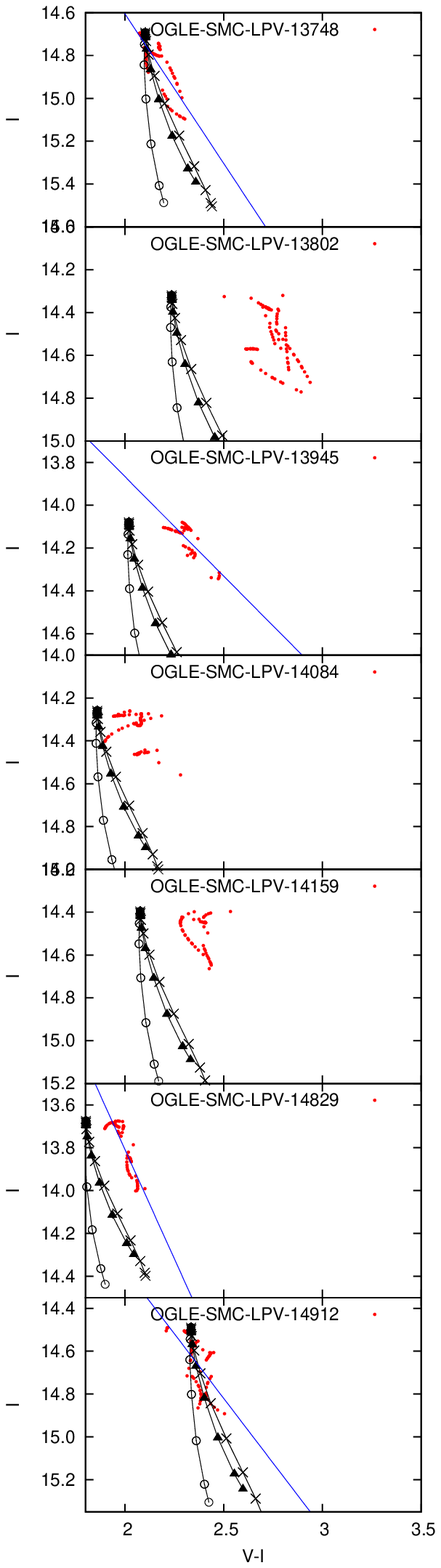}
\end{minipage}
\end{tabular}
\caption{
 The same as Figure \ref{spot_IJ} but using the $V-I$
  colour rather than $I-J$.
}
\label{spot_VI}
\end{figure*}

 In Figures \ref{spot_IJ}, \ref{spot_JK} and \ref{spot_VI} we show
the $I$ magnitude plotted against 
the colours $I-J$, $J-K_{s}$ and
$V-I$, respectively, for the observed
samples of oxygen-rich and carbon stars. The magnitudes of models at
the maximum light, which corresponds to when there is no visible 
spot, are adjusted in the plots to the magnitudes at the observed maximum light for
each star.  

As with our comparison of dust shell models with observations,
we test the spot models by looking for differences in the slopes in ($I$,colour) diagrams
of model lines and observational variations.  Models with
different ratios of the spot to normal photospheric temperature show different slopes.
The models with
higher spot temperature (i.e. larger temperature ratio) show shallower slopes.

For carbon stars, the slope of the line representing the observations generally
shows a good agreement with the model lines in each of 
Figures \ref{spot_IJ}, 
\ref{spot_JK} and \ref{spot_VI} (two exceptions are OGLE-SMC-LPV-13557 and
OGLE-SMC-LPV-13945 in Figure~\ref{spot_VI} which show disagreement with their
models). According to the figures, the models with hotter spots (temperature ratio 0.833)
and the hemisphere models (which also have a temperature ratio of 0.888) fit the models better
in each of the ($I$,$I-J$),
($I$,$J-K_{s}$) and ($I$,$V-I$) diagrams.  If sequence D light variations are due
to rotating spots, these results suggest that the spot temperatures in carbon
stars should be roughly 2500 K.

In distinct contrast to the carbon stars, the oxgen rich star variations do not agree with model
variations in any of the ($I$,$I-J$), ($I$,$J-K_{s}$) or ($I$,$V-I$) diagrams.
Although agreement would not be expected in the ($I$,$J-K_{s}$) plane where the observations are
affected by absorptions due to H$_{\rm 2}$O molecules, the bad disagreement in the
($I$,$I-J$) and ($I$,$V-I$) diagrams indicates that
spot models cannot explain the light and colour variations due to the LSP
phenomenon in oxygen-rich stars.  Since the LSPs of both oxygen-rich stars and carbon stars
should have the same origin, the failure of the spot models to reproduce
the observations in oxygen-rich stars means that spots are not a reasonable 
explanation for LSPs in sequence D stars. This result
agrees with the conclusion of \citet{oli03} that the rotation periods
of LSP stars are too long for a spotted star model to explain LSP variations.

\subsubsection{The effect on spot models of the inclination of the rotation axis and the spot lattitude}

\begin{figure}
\includegraphics[width=0.5\textwidth]{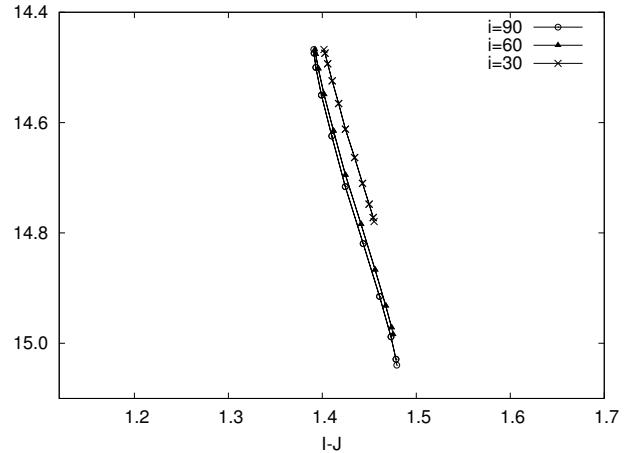}
\caption{The variation of $I$ with $I-J$ in 
hemisphere spot models for oxygen rich stars when different inclination angles for the
rotation axis are assumed.
}
\label{inc}
\end{figure}

The models above were made assuming spots on the equator of a rotating
star whose rotation axis is perpendicular to the line of sight.
Changing the lattitude of a spot will reduce its apparent area
and hence the amplitude of the light variation it will cause.
However, the slope of the locus of colour-magnitude variation should
be unchanged since the reduction in apparent area also occurs
as a spot on the equator rotates towards the limb of the star 
(the very minor effect of limb darkening will produce very
small differences between the two cases).  The same argument
applies in the case of a star with a tilted rotation axis.  We thus
expect the results of the previous section to apply regardless of the inclination
of the rotation axis and the lattitude of the spots.  As a test
calculation, we examined the change in the $I$ and $I-J$ magnitudes
for a star with the rotation axis tilted at 30, 60 and 90 degrees.
The results are shown in Figure~\ref{inc}.  It can be seen the
the slope of the variation is the same in all cases.

\subsection{A combined dust and dark spot model}

In Section~\ref{sec:comparison_dust} the possibility that a dust shell
model alone could explain the LSP phenomenon was examined and
rejected. Similarly, in Section~\ref{sec:comparison_spot} a rotating
spotted star model was examined and rejected.  Here we examine the
possibility that a combination of a rotating spotted star and dust
absorption at certain phases of the rotation could cause the observed
behaviour of stars with LSPs.

In this combined model, the effective temperature of the central star is
assumed to vary sinusoidally to simulate a rotating spotted star and
the optical depth of the dust shell is assumed to vary sinusoidally to
simulate periodic dust absorption.  The following equations are used:
\begin{eqnarray}
T_{\rm eff}(\phi)&=&T_{\rm eff,0}+\Delta T_{\rm eff}\sin \left( 2\pi \phi + \beta \right) \label{func_Teff}\\
\tau_{V}(\phi)&=&\tau_{V,0}\sin \left(\pi \phi \right)
\end{eqnarray}
where $\Delta T_{\rm eff}$ and $\tau_{V,0}$ are the amplitudes of
variation of the effective temperature and the 
optical depth in $V$ of the dust shell,
respectively.  $T_{\rm eff,0}$ is the average effective
temperature. 
As a direct consequence of these equations, the central star luminosity
is given by
\begin{eqnarray}
L(\phi)&=&\frac{T_{\rm eff}^4(\phi)}{T_{\rm eff,0}^4} L_{0}
\end{eqnarray}
where $L_{0}$ is the average luminosity. 
The fluxes emitted by the central star in the various photometric bands were
obtained assuming the central star was a blackbody.  The dust absorption
in photometric bands other than $V$ was obtained using \textsc{dusty}.  The
dust compositions used in Section~\ref{sec:model} for oxygen-rich and carbon-rich stars was adopted
and the shell outer radius was set to 1.5 times the inner radius, which roughly 
corresponds to the models described above with an expansion velocity of 15 km s$^{-1}$.
The inner dust shell temperature was set to 800\,K.

In the combined model, $\phi$ varies from 0 to 1 as the spotted star makes one complete
rotation.  The phase shift $\beta$ between the rotation phase and dust absorption phase 
has been introduced to allow for dust absorption at different possible rotation
phases.  Since we do not know how dust absorption might be linked to the rotation phase,
models were made with different values for $\beta$.
Note that the dust absorption goes through one maximum as the star makes
a full rotation.

Table~\ref{tab04} gives the parameters considered for the models described
here.  Two values of $\Delta T_{\rm eff}$, 50\,K and 100\,K, were used for both oxygen-rich and
carbon-rich stars.  The values of $\tau_{V,0}$ were adjusted so that the
variations in magnitude caused by dust dominated for the models with
$\Delta T_{\rm eff} = 50$\,K while the variations in magnitude caused by
$T_{\rm eff}$ variations dominated when $\Delta T_{\rm eff} = 100$\,K.

\begin{figure*}
\includegraphics[width=1\textwidth]{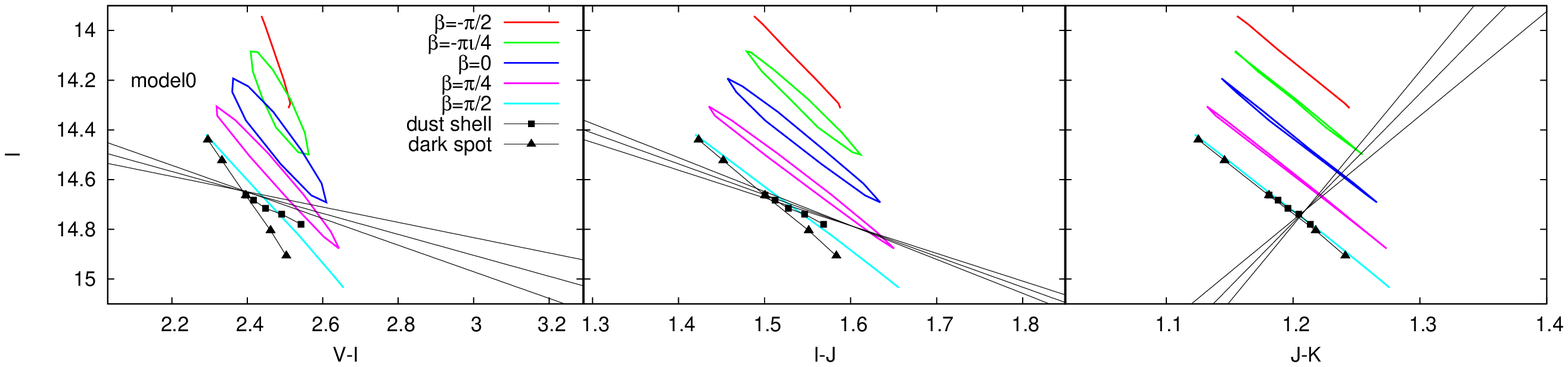}\\
\includegraphics[width=1\textwidth]{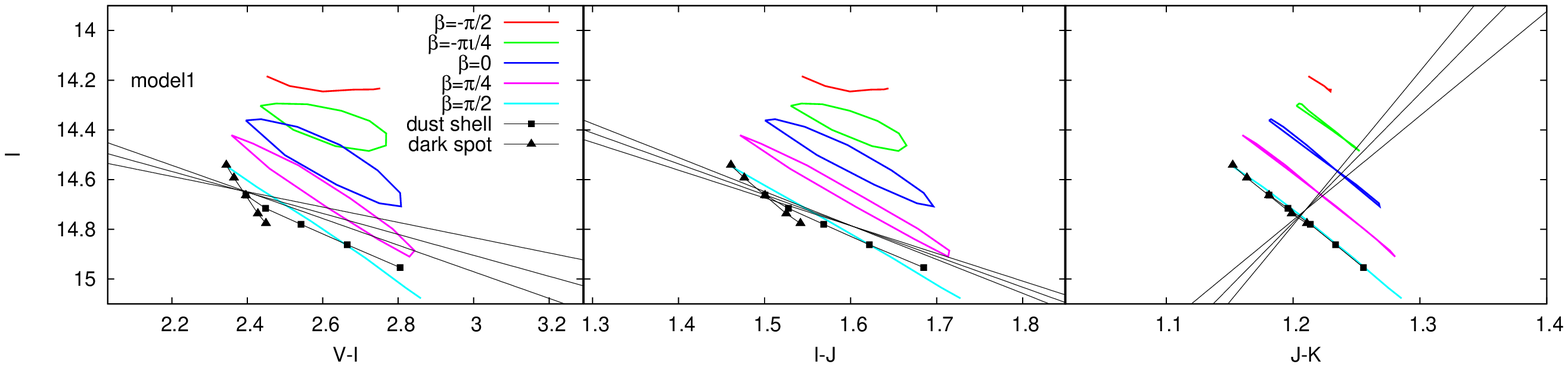}\\
\includegraphics[width=1\textwidth]{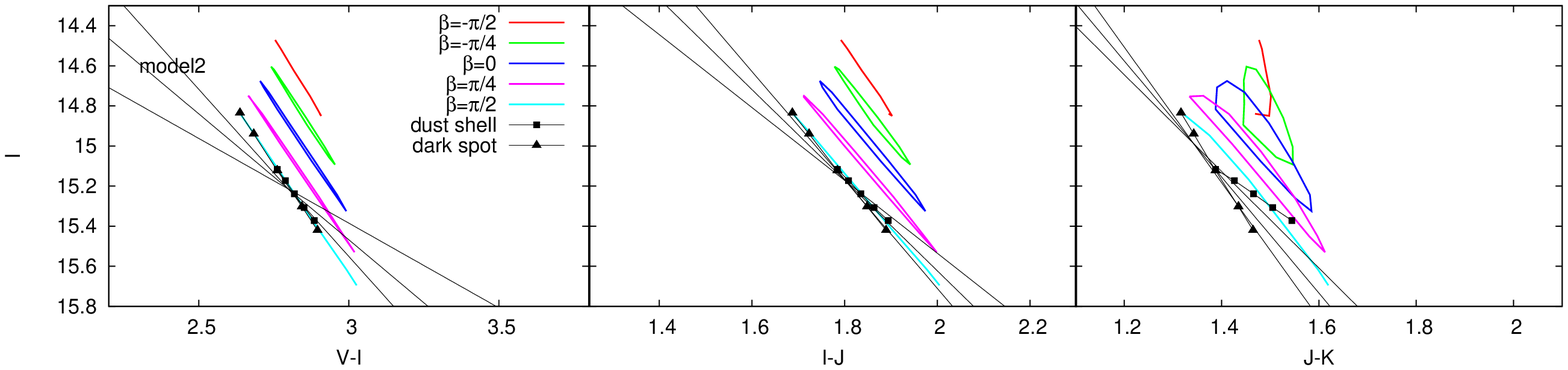}\\
\includegraphics[width=1\textwidth]{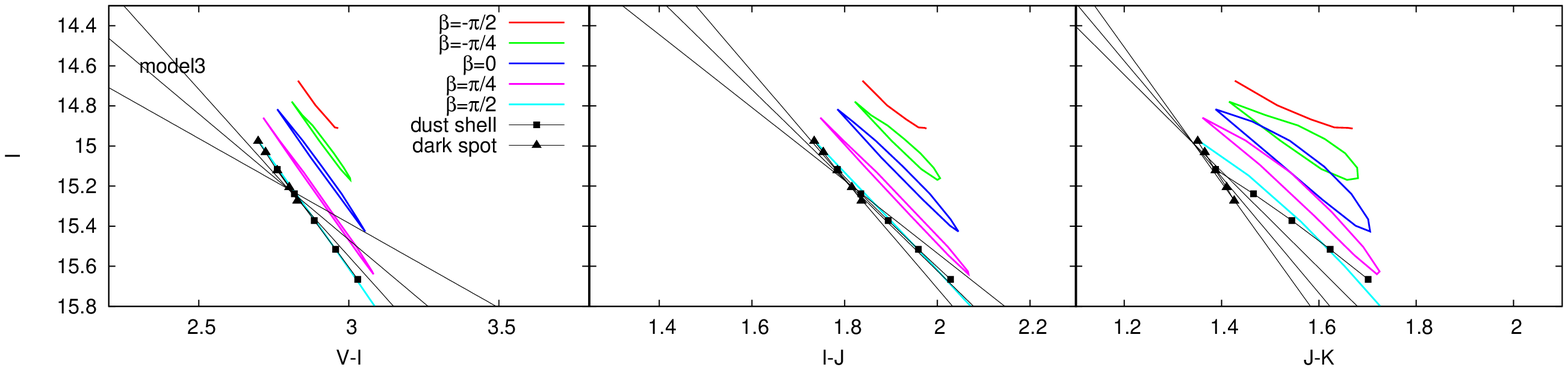}
\caption{
The variations of $I$ with $V-I$, $I-J$ and $J-K$
for the models listed in Table~\ref{tab04}. Models for oxygen-rich stars are
shown in the upper 2 panels and models for carbon-rich
stars are shown in the lower 2 panels.  Solid lines with filled
squares represent the dust shell model alone while solid lines
with filled triangles represent the dark spot model alone. 
The combined dust and rotating spotted star models are shown as coloured lines with the colour
indicating the phase lags $\beta$.  Note that the coloured lines have
been arbitrarily shifted in $I$ to prevent confusing overlap since
it is only the slope of the lines that is of interest.  The slopes of
the three solid black lines in each diagram represent the observed variations.
The central line corresponds to the mean value of the observed line slope 
for stars with a relative error in the slope of
less than 0.2 and the outer lines vary from the mean slope by 1
$\sigma$. 
}
\label{comb_model}
\end{figure*}

Figure~\ref{comb_model} shows the $I$ magnitude plotted against $V-I$, 
$I-J$ and $J-K$ for the various models computed here.  In general,
the slopes of the variations in the ($I$,colour) diagrams for the combined
models lie between the slopes for pure dust absorption and pure
$T_{\rm eff}$ variations (spot model).  The combined models do not
generally represent an improvement over the pure dust or spot models.
In particular, they can not explain the observation that $J-K$ in oxygen-rich stars
get bluer as the star dims.

\begin{table}
\begin{center}
\caption{$\Delta T_{\rm eff}$ and $\tau_{V, 0}$ for models}
\begin{tabular}{ccccc}
\hline
\hline
Sp. type& $T_{\rm eff,0}(K)$ & $\Delta T_{\rm eff}$(K) & $\tau_{V, 0}$ & model\\
\hline
O rich star & 2800 &100 & 1 & model0 \\
& 2800 & 50 & 2 & model1\\
\hline
C rich star & 2500 & 100 & 0.5 & model2\\
& 2500 & 50 & 1 & model3\\
\hline
\end{tabular}
\label{tab04}
\end{center}
\end{table}

\section{Summary}

We have examined broadband observations of samples of oxygen-rich and carbon stars
in the SMC which exhibit LSP light variations.  The observations are in the $VIJHK_s$ bands 
and they were obtained by the OGLE project
and by the SIRIUS camera of the IRSF 1.4m telescope.

The $J$ and $K_s$ observations
of oxygen-rich stars reveal a significant new feature of LSP light variations.
As light declines, the $J-K_s$ colour barely changes or sometimes gets bluer.
We interpret this as indicating that cool gas containing a significant amount of
H$_{\rm 2}$O is levitated above the photosphere of LSP stars at the beginning of
light decline.  Absorption of light in the $K_s$ band by H$_{\rm 2}$O forces the $J-K_s$
colour to remain constant or to get bluer during light decline.  The levitation of
matter above the photosphere found here, along with the discovery of
the development of a warm chromosphere during rising light by \citet{woo04}, provides strong 
evidence that the LSP phenomenon is associated with mass ejection
from the photosphere of red giant stars.

We examined two possible models for the explanation of the LSP
phenomenon by comparing these models with broadband 
observations of both oxygen-rich and
carbon stars.  The first model 
assumes LSP light variations are
due to dust absorption by ejected dust shells
while the second model assumes a rotating star with a dark spot.

In the dust shell ejection model for LSP stars, it was found that dust absorption by
the ejected mass shell could not explain the
observed variations in the ($I$,$J-K_{s}$) or
($I$,$V-I$) diagrams for either oxygen-rich or carbon stars.  
In oxygen-rich stars, the models fail to reproduce the variations
in the ($I$,$J-K_{s}$) because the models only include variable
dust absorption and not the variable H$_{\rm 2}$O absorption noted above.  
The failure of the dust shell models to reproduce
the observed variations in the ($I$,$V-I$) diagram for either oxygen-rich or carbon stars,
and the failure to reproduce the observed variations in the ($I$,$J-K_{s}$) diagram
for carbon stars, suggests that dust absorption in ejected mass shells is
not the cause of LSP light variations.

The LSP model involving a rotating, spotted star was not able to reproduce 
the observed variations in oxygen-rich stars in either the ($I$,$I-J$)
or ($I$,$V-I$) diagrams (or the $I$,$J-K_s$ diagram, but here the 
$J-K_s$ colour is affected by H$_{\rm 2}$O absorption that is not included in
the models).  We thus conclude that the rotating, spotted star model is
not the explanation for LSP light variations either.  This result
strengthens the conclusion of \citet{oli03} that the rotation periods
of LSP stars are too long for a spotted star model to explain LSP variations.
A simple combined dust and spotted star model was no
better at explaining the observed magnitude-colour variations than the
dust and spotted star models individually.

Our conclusion is that some process other than dust shell ejection or
spot rotation lies behind the light variations seen in stars with LSPs.
This unknown process seems to be responsible for the ejection of matter
near the beginning of light decline.  The ejected matter may be
the origin of the excess mid-infrared emission due to dust that is seen
in stars with LSPs \citep{woo09}.

\section*{acknowledgements}

M.T. gratefully acknowledges the hospitality and generous support of Prof.
Peter Wood and his colleagues during his long term visit at the
Australian National University where this work was written. He would also
like to thank Prof. Hideyuki Saio for useful discussions and help.
The authors thank the referee for constructive comments.
This research is supported by Brain Circulation Program (R2301) of the
Ministry of Education, Science, Culture, and Sports in Japan. This
research is partly supported by the Japan Society for the Promotion of Science through Grant- in-Aid for Scientific Research 26$\cdot$5091.

\bsp

\label{lastpage}

\end{document}